# Theory of the charge response of the boron-vacancy spin in monolayer hexagonal boron nitride

Dongyeon (Tyler) Kim[1], Theodore Scharlotta[2], and Daniel Hashemi[1*]

[1] Department of Physics, Optical Engineering, and NanoEngineering, Rose-Hulman Institute of Technology, 5500 Wabash Avenue, Terre Haute, Indiana 47803, USA

[2] Department of Chemical Engineering, Rose-Hulman Institute of Technology, 5500 Wabash Avenue, Terre Haute, Indiana 47803, USA

*Contact author: hashemi@rose-hulman.edu

ORCID: Dongyeon (Tyler) Kim 0009-0000-4616-2695; Theodore Scharlotta 0009-0007-7485-717X; Daniel Hashemi 0000-0001-7328-1032

## Abstract

The boron vacancy $V_{\mathrm{B}}^{-}$ in hexagonal boron nitride is a leading two-dimensional spin sensor, but the microscopic connection between nearby charge positions and its spin Hamiltonian remains poorly understood. Calculations for carbon and oxygen donors 3.85–8.08 Å away show that direction matters, not only distance. In the frozen lattice, reversing a carbon donor's azimuth at 4.35 Å (C1 to C1m) more than halves $|E|$ and turns its axis across the mirror line. The ratio is 0.467 with the Perdew–Burke–Ernzerhof (PBE) functional, against 1 for distance alone. A field-plus-gradient model fitted to six donor configurations, including C1 but excluding C1m, predicted a ratio of 0.443 before C1m was calculated. Heyd–Scuseria–Ernzerhof hybrid statics give 0.432 and PBE relaxation 0.600. Exploratory analysis suggests why direction matters: a charge's potential differs at the three vacancy nitrogens and moves the spin of the two $e'$ holes among them, more when the charge faces one nitrogen than the gap between two. $E$ follows this spin imbalance through one constant, $c \approx 8$ GHz per $\mu_{\mathrm{B}}$, across donors and applied fields, whereas in PBE $D$ is nearly insensitive to it. A three-site model with one fitted screening constant reproduces the frozen-lattice $E$ to 83 MHz rms. Results are for one charge state in a free-standing monolayer; absolute $|E|$ varies by tens of percent across approximations.

## 1. Introduction

The boron vacancy $V_{\mathrm{B}}^{-}$ in hexagonal boron nitride (hBN) is a spin-triplet defect read out optically at room temperature [1] and the reference spin sensor of van der Waals materials [2,3]. It has sensed magnetic fields, temperature and pressure [4] and imaged magnetism [5,6]. In few-layer flakes [7] a sensor nanometers from a sample also sits nanometers from charged defects. These charges shift its transition frequencies by a static offset that must be calibrated and that, used in reverse, could constrain where a charge sits. Both uses require knowing what a charge does to the spin, and why.

The vacancy responds to charges because two triplet-coupled holes [8] carry its spin, mainly on the three neighboring nitrogens (Fig. 1(a)) [9,10]. Udvarhelyi et al. computed that strain acts mainly on the axial zero-field parameter $D$ and an in-plane field mainly on the transverse parameter $E$, which they attributed to the fields of charged defects [11]. In field-only point-charge models [12,13], for isotropically screened in-plane charges, distance alone sets $|E|$, and direction only its axis. Two of us calculated that the carbon–vacancy complex $V_{\mathrm{B}}C_{\mathrm{B}}^{0}$ couples strongly to strain, while its spin–spin tensor shows no resolved linear clamped-ion response to in-plane fields [14]. These studies leave open how a donor's direction enters at fixed distance and which electronic mechanism carries a charge's potential into the spin tensor.

Here we show that a charge at a fixed distance from the vacancy can produce a substantially different transverse splitting depending on its direction, likely because it redistributes spin among the three nitrogens. Reversing the azimuth of a carbon donor 4.35 Å from the vacancy (C1 to C1m) halves $|E|$ in the frozen lattice and turns its axis across the donor's mirror line. $E$ follows one quantity, the spin imbalance around the nitrogen triangle, whereas with the Perdew–Burke–Ernzerhof (PBE) functional $D$ is nearly insensitive to it. The direction dependence is the tested result of this work; the spin picture is an exploratory explanation of it.

Defect-chemistry calculations single out carbon and oxygen as donors in hBN [15], and carbon paired with the vacancy is a prototype $S = 1$ donor–acceptor system [16] with computed $D$ [17]. We therefore place a $C_{\mathrm{B}}^{+}$ or $O_{\mathrm{N}}^{+}$ donor 3.85–8.08 Å from an intact vacancy. With PBE we compute eight placements before and after relaxation, C1–C4 and O1–O4, numbered outward (Fig. 1(b)), and three angular controls that reverse the donor azimuth $\psi$. A screened field-plus-gradient model fitted to six calibration placements, C1 among them, predicted two reserved placements and the controls before their tensors were computed. We treat one charge state, a neutral complex of $V_{\mathrm{B}}^{-}$ and an ionized donor, in a free-standing monolayer, and do not compute which charge states and placements a sample realizes.

## 2. Theory and computational approach

We describe each configuration as one electronic spin $S = 1$ with a spin–spin zero-field tensor $\mathbf{D}$, coupled to the three vacancy-neighbor $^{14}$N nuclei (Sec. S1 of the Supplemental Material [18]), and neglect spin–orbit coupling. With $x$ along the lattice vector $\mathbf{a}_1$ and $z$ along the sheet normal,

$$\begin{aligned} D &= \frac{3}{2} D_{zz}, \\ E_+ \equiv E_1 + \mathrm{i}E_2 &= \frac{D_{xx} - D_{yy}}{2} + \mathrm{i}\, D_{xy} = |E| \exp(2\mathrm{i}\theta_E)\,. \end{aligned} \tag{1}$$

$D$ measures the zero-field splitting of the triplet, and $\Delta D = D - D(V_{\mathrm{B}}^{-})$ its change by a donor. $|E|$ measures how strongly a donor breaks the threefold symmetry, and $\theta_E$ is the direction from $\mathbf{a}_1$ of the $E$ axis, the in-plane principal axis with the larger eigenvalue. We keep $z$ along the normal even where $|E| > D/3$ (C1 and O1). The vacancy nitrogens N30, N150 and N270 lie at azimuths $\beta_j$ = 30°, 150° and 270° from $\mathbf{a}_1$ about the vacancy center in the ideal lattice, each on a mirror line. For a donor on such a line, symmetry fixes the $E$ axis along or perpendicular to it (the two branches).

Carbon replaces a boron and oxygen a nitrogen atom, never a vacancy neighbor (Fig. 1(b)). The eight placements lie in the sector $0° \leq \psi \leq 30°$, where $\sin 3\psi \geq 0$. The three angular controls reverse that sign: C1m and C3m, at the C1 and C3 distances, and the oxygen donor Om at 5.24 Å. We evaluate each placement after relaxation ($G_X$) and in the frozen lattice ($G_0$): the relaxed $V_{\mathrm{B}}^{-}$ with the donor substituted, no atom moved and the electrons self-consistent (Fig. 1(c)). The two geometries separate frozen-lattice from relaxation contributions, not electrostatics from chemistry.

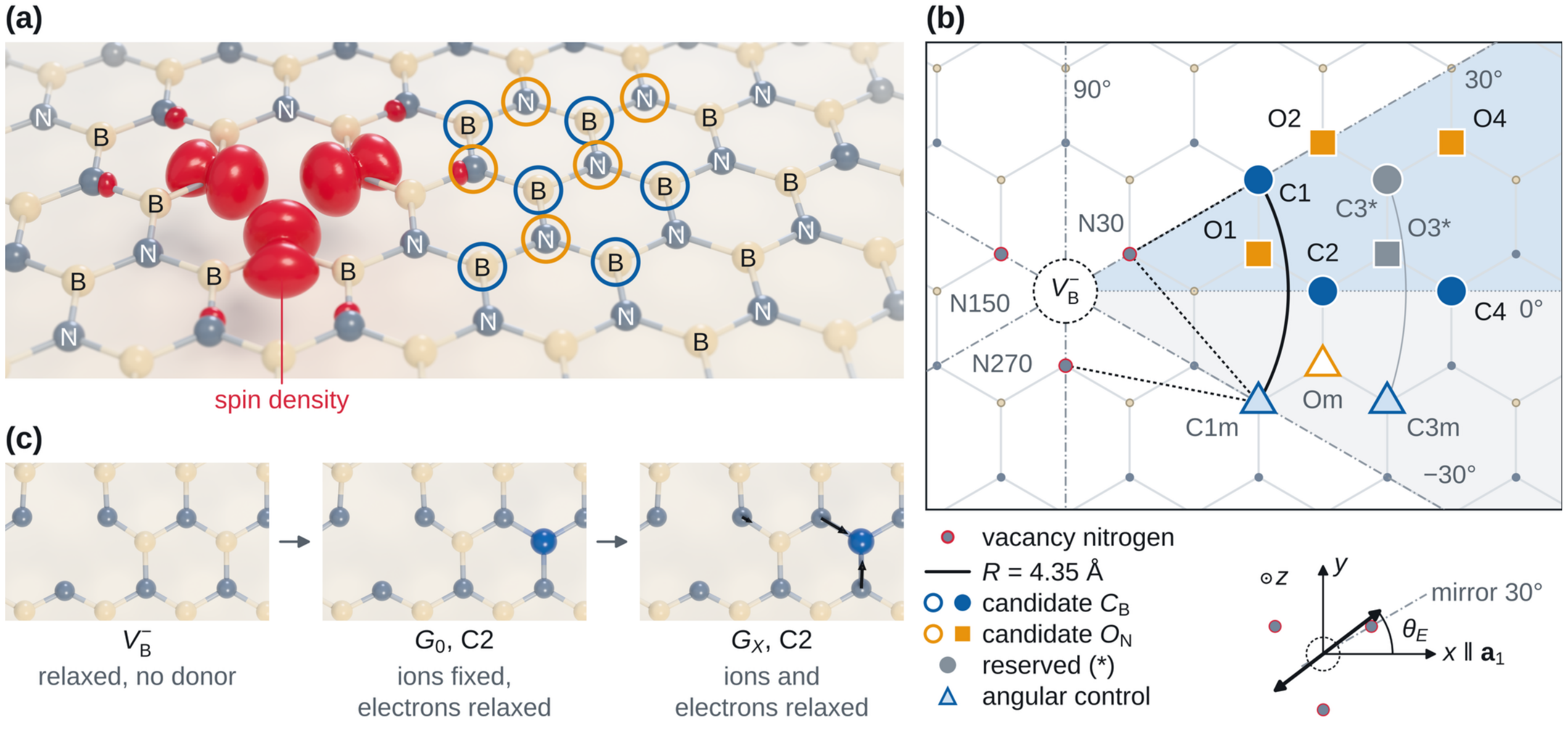

**Figure 1.** Donor sites and the vacancy's spin (PBE, $\mathbf{8\times 8}$ cell). (a) Spin density of the relaxed $\boldsymbol{V_B^-}$ alone, isosurface $+\mathbf{0.06}\,\boldsymbol{\mu}_\mathbf{B}\,\text{Å}^{-3}$ (crimson; smooth part), on the three vacancy nitrogens. Boron cream, nitrogen slate in every figure. Rings mark the eleven candidate donor sites ($\boldsymbol{C_B}$ azure, $\boldsymbol{O_N}$ saffron), each a separate one-donor calculation. (b) Donor sites and dash-dot mirror lines. Arcs join equal-distance placements; the eight placements fill the blue 0–30° sector. C1 and C1m, both at 4.35 Å, lie on mirror lines, but the reflection between them is not a symmetry of the vacancy: C1 has one nearest vacancy nitrogen, C1m two (dotted lines). (c) $\boldsymbol{G_0}$, donor in the relaxed $\boldsymbol{V_B^-}$, ions fixed, electrons self-consistent; $\boldsymbol{G_X}$, ions also relaxed (C2; displacements magnified 12 times). Later figures draw $\boldsymbol{G_0}$ open and $\boldsymbol{G_X}$ filled. Lower right, lab frame and the angle $\boldsymbol{\theta_E}$ of an $\boldsymbol{E}$ axis.

We use density-functional theory (DFT) with PBE [19] in the projector-augmented-wave (PAW) code VASP [20,21], with a truncated Coulomb kernel [22] and the total spin fixed at $S_z = 1$ (Sec. S2). Unless marked, values are uncorrected PBE in an $8 \times 8$ supercell at the $\Gamma$ point. We report the $\mathbf{D}$ of the unrestricted Kohn–Sham determinant [23,24]: $D$ is 2798.8 MHz for $V_B^-$, about 20% below the 3.48 GHz measured at 300 K [1], as in earlier PBE monolayer calculations [25]. Donor-induced differences cancel only the error shared with the bare vacancy, and $|E|$, zero there by symmetry, carries its method error directly. Every PBE tensor passed an electronic-state check (Sec. S3), though we claim no global $S = 1$ ground state.

We distinguish three sources of uncertainty: numerical convergence, finite-size effects and the electronic-structure approximation. Numerical uncertainty is at most 0.28 MHz in $\Delta D$ and 2.40 MHz in $|E|$ for the calibration placements (Sec. S6). Periodic images change $|E|$ far more than $D$: from $8 \times 8$ to $10 \times 10$ the relaxed $|E|$ rises by 7.3% for the near donor C2 and 35.3% for the far donor O4 (Sec. S5). The level of theory matters most (Sec. S8). Heyd–Scuseria–Ernzerhof (HSE06) hybrid-functional statics [26,27] at PBE geometries raise the $|E|$ of C2 and O2 by 54–56%, and spin decontamination [28] lowers it, post hoc, by 37–53%. Neither test bounds it, so $|E|$ is a PBE value, whereas ratios at fixed distance vary less (Sec. 4).

For tests fixed in advance, we recorded the criteria before reading the results (Sec. S9). The $x$ and $\xi$ relations of Sec. 4, the analyses of Secs. 3 and 5 and the map, ensemble and noise estimates of Sec. 6 are exploratory, chosen after all those results were known. Post hoc marks any other check made after those results were known.

## 3. Electronic structure of the vacancy

The spin of $V_B^-$ is carried by two holes in its in-plane $e'$ doublet. Removing a boron atom leaves three nitrogens whose in-plane dangling and out-of-plane $p_z$ orbitals combine in $D_{3h}$ symmetry into in-plane $a_1'$ and $e'$ and out-of-plane $a_2''$ and $e''$ levels [8,9]. The pristine monolayer's PBE gap is 4.66 eV at K, against a 6.1 eV optical gap measured on epitaxial monolayers [29]. Unfolded onto the primitive cell [30], the spectral function of the 8 × 8 cell keeps the host bands and adds flat, spin-split vacancy levels in the gap (Fig. 2(a)). Every vacancy level is filled except the minority $e'$ doublet, 2.35 eV above the valence-band maximum and 2.21 eV below the conduction band (Fig. 2(c)). Its two holes form the triplet ground state $^3A_2'$ [8,10], and the exchange splitting of $e'$, 1.65 eV at Γ, keeps them there. Projected densities of states (Fig. 2(b)) and orbitals (Fig. 2(d)) place them on in-plane $p$ orbitals of the three nitrogens.

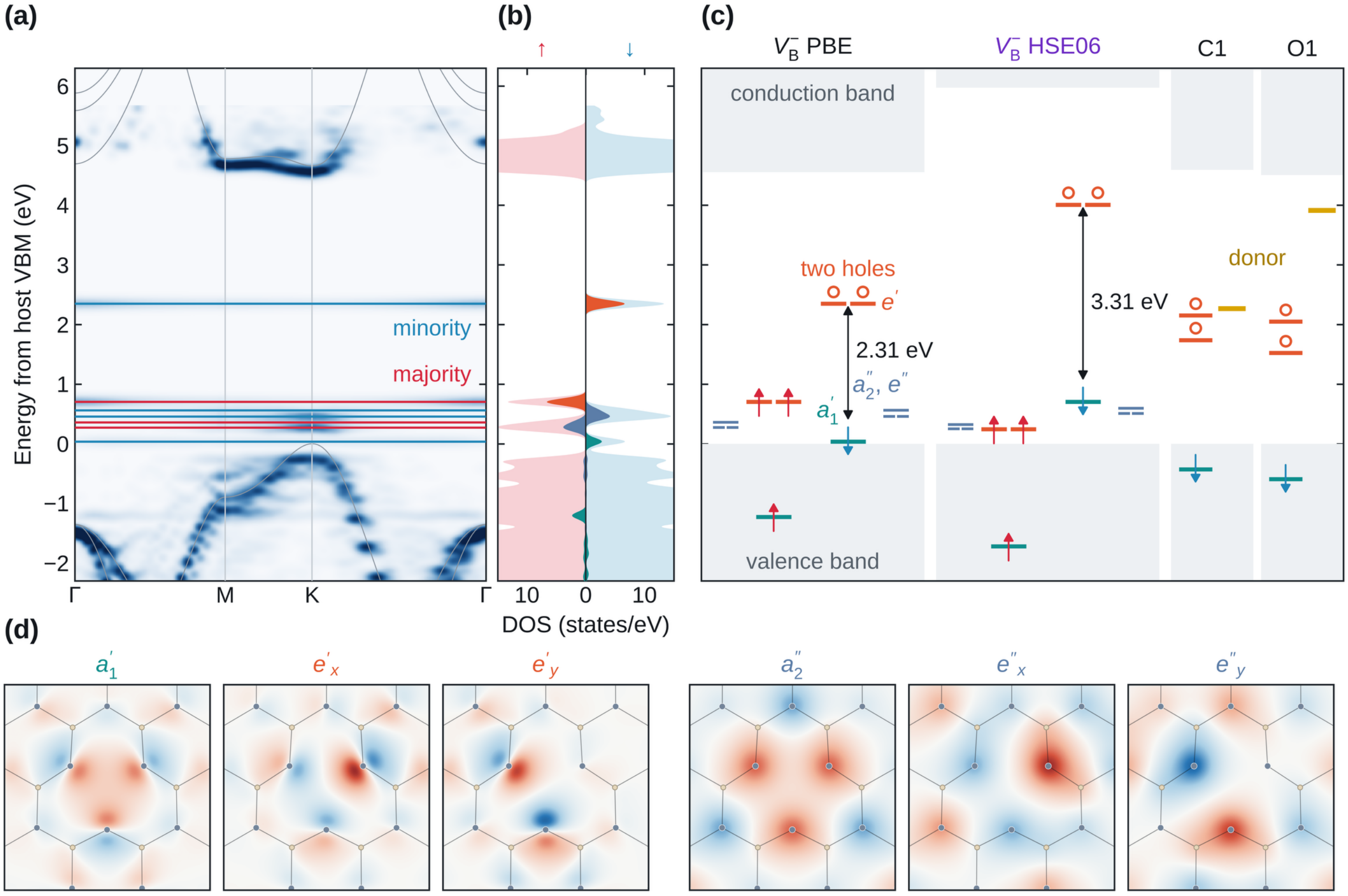

**Figure 2.** Electronic structure of $V_B^-$ and the levels that carry its spin (PBE unless marked). (a) Spectral weight of the $8 \times 8$ cell unfolded onto the primitive cell, pristine-sheet bands as hairlines, and the vacancy levels of the majority (crimson) and minority (cerulean) spin. (b) Spin-resolved density of states (tint) and its projection on the three vacancy nitrogens: $a_1'$ (teal), $e'$ (coral), $p_z$ (slate blue). (c) Γ-point levels with $D_{3h}$ labels, PBE and HSE06: arrows, electrons; circles, holes; numbers, the minority $a_1'$–$e'$ gap. C1 and O1 ($G_0$, PBE, minority spin) split $e'$ and add an empty donor level (gold). (d) Minority orbitals at Γ: $a_1'$ and $e'$ in the sheet plane, $a_2''$ and $e''$ 0.75 Å above it; red and blue, the signs of the wave function. Energies from the host valence-band maximum of each calculation.

In PBE one energy governs what follows: the minority gap Δ from the filled $a_1'$ level to $e'$, 2.315 eV at Γ. It stays within 2.27–2.41 eV in the eleven complexes, frozen or relaxed, so a donor perturbs this level scheme without rebuilding it (Sec. S11). The donor is ionized: its levels are empty, with carbon's lowest close to $e'$ and oxygen's well above it, and its moment stays below 0.02 $\mu_B$ for the calibration placements (Sec. S3, Table S1). These energies are Kohn–Sham differences at Γ, not charge-transition levels [31] or optical thresholds [10]. HSE06 statics reorder several levels and widen these gaps: the vacancy cell's host gap rises from 4.56 to 5.97 eV and Δ to 3.31 eV.

## 4. Direction, distance and relaxation

At fixed distance the azimuth changes both $|E|$ and its axis (Fig. 3, Table I). The clearest case is the C1/C1m pair at 4.35 Å. C1, at azimuth +30°, faces N30; C1m, at −30°, faces the midpoint between N30 and N270. Reversing the azimuth lowers the frozen-lattice $|E|$ from 987.6 to 461.5 MHz (ratio 0.467, Fig. 3(a)), and the axis switches branch, from along the donor's mirror line ($\theta_E$ = 30.0°) to across it (60.0°, Fig. 3(c)). Any description in vacancy-center distance alone, the field-only point-charge model included, gives 1, whereas the field-plus-gradient model of Sec. 6, with C1 in its fit, gave 0.443 before C1m was computed. C3 and C3m, at 6.65 Å, repeat the pattern with frozen-lattice ratio 0.600.

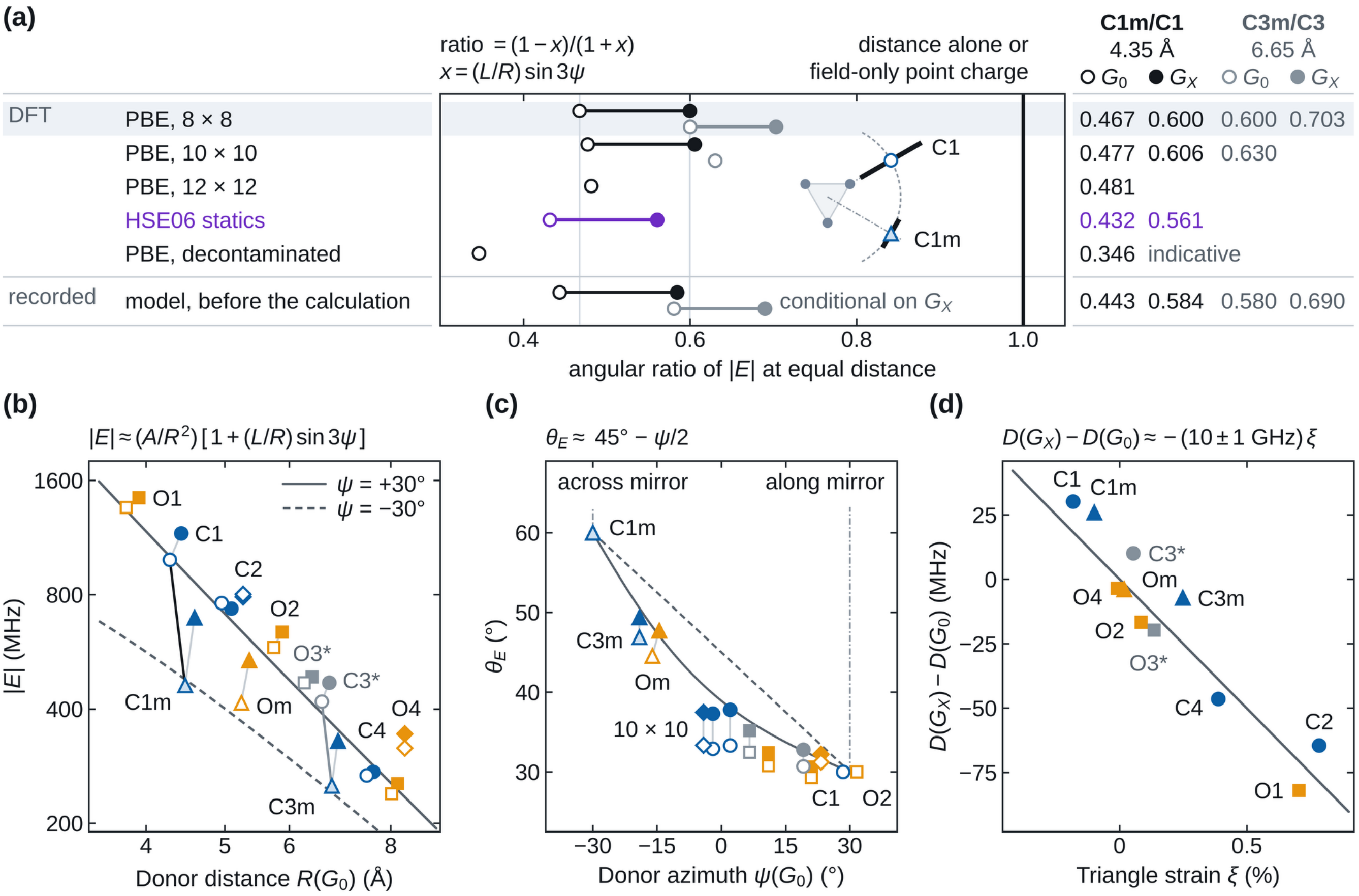

**Figure 3.** Direction and distance dependence of the tensor. Uncorrected PBE, $\mathbf{8\times 8}$ cell, PBE geometries, unless marked. (a) Ratio of $|\boldsymbol{E}|$ at equal distance, C1m/C1 (upper, black) and C3m/C3 (lower, gray): DFT and the model prediction recorded beforehand (Sec. S9), its $\boldsymbol{G_X}$ values given relaxed structures. Open $\boldsymbol{G_0}$, filled $\boldsymbol{G_X}$. Inset: C1 and C1m; bars, $\boldsymbol{E}$ axes; formula, Sec. 4. (b) $|\boldsymbol{E}|$ against $\boldsymbol{R}$; lines join equal-distance partners; curves, the far-field expression of Sec. 6, effective $\boldsymbol{L}$ = 1.4 Å from the isolated-complex ratios of Sec. S5. (c) $\boldsymbol{\theta_E}$ against azimuth $\boldsymbol{\psi}$; dashed, far-field expression; solid, Eq. (5) at $\boldsymbol{R}$ = 5 Å. (d) Relaxation step of $\boldsymbol{D}$ against triangle strain $\boldsymbol{\xi}$; line, the fit of Sec. 4. In (b) and (c) diamonds are the $\mathbf{10\times 10}$ cell; saffron triangles Om, gray asterisked symbols C3 and O3; legibility offsets ≤ 0.24 Å and 4.2°.

The contrast persists in larger cells, with a hybrid functional and after relaxation (Fig. 3(a)). Frozen-lattice C1m/C1 is 0.477 and 0.481 in 10 × 10 and 12 × 12 cells, under criteria fixed in advance (Sec. S5), and 0.432 in HSE06 statics, 7.6% below PBE and outside the ±5% fixed beforehand for agreement. Relaxation raises C1m/C1 to 0.600 and C3m/C3 to 0.703, against 0.584 and 0.690 predicted given the relaxed structures. Spin decontamination of the frozen-lattice tensors gives 0.346, an indication, not a correction (Sec. S8). The C1m axis keeps its branch throughout. All these ratios lie below 1: they support a substantial angular reduction, not one precise ratio.

**Table I.** Donor-induced changes of the $V_{\mathrm{B}}^{-}$ spin Hamiltonian. Uncorrected PBE, $\mathbf{8}\times\mathbf{8}$ cell, MHz unless noted. Paired entries are $\boldsymbol{G_0}$ / $\boldsymbol{G_X}$. $\boldsymbol{R}$, $\boldsymbol{\psi}$: donor distance and azimuth from $\mathbf{a_1}$ at $\boldsymbol{G_0}$; C1/C1m and C3/C3m share a distance, not an azimuth. $\boldsymbol{\Delta D = D - D(V_{\mathrm{B}}^{-})}$; differences use unrounded values. $\boldsymbol{\Delta A_{zz}}$: change of the N30 $^{\mathbf{14}}\mathbf{N}$ hyperfine component along the normal, with core polarization (absolute for $\boldsymbol{V_{\mathrm{B}}^{-}}$). Reserved (*) and angular-control rows enter no fit of Eq. (6). $|\boldsymbol{E}|$ is a PBE value (Sec. 2).

| **Donor** | $R$ **(Å)** | $\psi$ **(°)** | $D$ | $\Delta D$ | $\lvert E\rvert$ | $\theta_E$ **(°)** | $\Delta A_{zz}$ **(N30)** |
|---|---|---|---|---|---|---|---|
| $V_{\mathrm{B}}^{-}$ (no donor) | – | – | 2798.8 | 0.0 | 0.006 | – | (34.99) |
| *Calibration placements (used to fit the model)* | | | | | | | |
| C1 | 4.35 | 30.0 | 2718.4 / 2748.6 | −80.5 / −50.3 | 987.6 / 1160.5 | 30.0 / 30.0 | +3.01 / +4.52 |
| C2 | 5.02 | 0.0 | 2744.7 / 2680.1 | −54.1 / −118.7 | 759.9 / 735.7 | 32.9 / 37.3 | +3.00 / +0.07 |
| C4 | 7.54 | 0.0 | 2781.2 / 2734.7 | −17.6 / −64.1 | 267.4 / 273.2 | 33.3 / 37.8 | +1.51 / +0.44 |
| O1 | 3.85 | 10.9 | 2782.7 / 2700.8 | −16.1 / −98.0 | 1357.4 / 1439.4 | 30.8 / 32.4 | +4.31 / +0.26 |
| O2 | 5.81 | 30.0 | 2782.9 / 2766.3 | −15.9 / −32.6 | 581.0 / 637.6 | 30.0 / 30.0 | +2.48 / +2.27 |
| O4 | 8.08 | 21.1 | 2783.5 / 2780.0 | −15.3 / −18.8 | 239.3 / 254.4 | 29.3 / 30.6 | +1.45 / +1.26 |
| *Reserved (predicted before they were computed)* | | | | | | | |
| C3* | 6.65 | 19.1 | 2747.1 / 2757.3 | −51.7 / −41.5 | 418.1 / 468.8 | 30.7 / 32.8 | +1.95 / +2.02 |
| O3* | 6.33 | 6.6 | 2782.1 / 2762.4 | −16.7 / −36.5 | 468.6 / 486.1 | 32.4 / 35.2 | +2.29 / +1.74 |
| *Angular controls (reversed azimuth)* | | | | | | | |
| C1m | 4.35 | −30.0 | 2720.4 / 2746.4 | −78.4 / −52.5 | 461.5 / 696.0 | 60.0 / 60.0 | −0.97[a] / +3.08[a] |
| C3m | 6.65 | −19.1 | 2750.7 / 2743.4 | −48.2 / −55.4 | 250.9 / 329.6 | 46.9 / 49.4 | +0.69 / +1.01 |
| Om | 5.24 | −16.1 | 2771.8 / 2767.8 | −27.1 / −31.0 | 415.2 / 536.9 | 44.5 / 47.7 | +0.88 / +2.36 |

[a] Same on N270.

Write each ratio as $\rho = (1-x)/(1+x)$, with $x$ the half-difference of the two $|E|$ over their mean. The far-field expression of Sec. 6 gives $x = (L/R)\,\sin\,3\psi$, so a pair that lies farther out, or has a smaller $\sin\,3\psi$, has a ratio closer to 1. That is the trend from C1 to C3. The effective length that fits the frozen-lattice C1/C1m pair, $L$ = 1.58 Å, even overshoots it, giving C3m/C3 = 0.67; the C3/C3m pair itself needs 1.97 Å. Relaxation multiplies $x$ by about 0.7 in both pairs.

Distance still orders the eight placements of this sector: $|E|$ falls monotonically with $R$, with carbon and oxygen interleaved, and the frozen lattice carries at least 85% of the relaxed value (Fig. 3(b)). Outside it that order fails: C1m, as close as C1, has nearly the frozen-lattice $|E|$ of O3 at 6.33 Å. The $E$ axis stays within 7.8° of the mirror line through N30 at all eight placements, but not at the controls: at C3m the frozen-lattice $\theta_E$ is 46.9°.

The axial parameter behaves differently. In the frozen lattice the carbon shift falls steeply with distance, whereas that of O1–O4 stays flat (Table I), but this contrast compares different sublattices and depends, post hoc, on spin treatment and cell size (Secs. S5 and S8). Reversing a carbon donor's azimuth barely moves the shift in PBE (C1m minus C1, +2.0 MHz), though it does in HSE06 statics (−16.0 MHz). Relaxation then moves $D$ in either direction, by up to 82 MHz (O1), and this step follows the size of the nitrogen triangle (Fig. 3(d)): $D(G_X) - D(G_0) \approx -(10 \pm 1\ \text{GHz})\,\xi$, with $\xi$ the relative change of the mean vacancy-nitrogen spacing ($r$ = −0.95, 11.5 MHz rms, carbon above oxygen), against −8.4 GHz for dipolar $d^{-3}$ scaling. We decomposed only C2 and O2: in geometry-only calculations of the bare vacancy, the lattice distortion alone gives 83% and 88% of the relaxed $\Delta D$ but at most 13% of the relaxed $|E|$ (Sec. S7), as in the strain–field split [11].

## 5. How a charge acts on the spin

We work back from the tensor to the charge: $E_+$ follows a spin imbalance through one constant, the charge's potential moves the two holes among the nitrogens, and the triangle's geometry makes that move depend on direction.

**The hole-pair tensor.** The zero-field tensor of a triplet is the dipolar interaction of its unpaired spins, so we evaluated it for the two holes alone, with pseudo-orbitals rescaled to unit norm (Sec. S11). For $V_B^-$ the pair gives $D$ = 2817 MHz, within 1% of the full value: $D$ is the dipolar energy of two holes spread over three nitrogens.

In a minimal model this picture can be solved exactly. Take one in-plane orbital $|j\rangle$ on each nitrogen and let the filled minority level be $|a\rangle = \sum_j a_j \, |j\rangle$. The holes span the two states orthogonal to $|a\rangle$, so the hole density on nitrogen $j$ is $n_j = 1 - a_j^2$, and the pair sits on the triangle edge opposite $j$ with probability $a_j^2$. That edge contributes a dipolar tensor with its axis along the edge, perpendicular to $\beta_j$. Summing the three edges gives

$$D = D_\mathrm{h} \sum_j a_j^2 = D_\mathrm{h}\,, \qquad E_+ = -D_\mathrm{h} \sum_j \delta\, n_j \exp(2\mathrm{i}\beta_j)\,, \tag{2}$$

where $\delta n_j = n_j - 2/3$ and $D_\mathrm{h}$ is the $D$ of the hole pair in the symmetric vacancy. In this picture, moving holes around the triangle leaves $D$ unchanged at every order, because the $a_j^2$ sum to one, but changes $E_+$ linearly, through a phase that turns twice as fast as the nitrogen azimuth. Direct evaluation of the pair supports both results. Over the 22 Γ-point tensors its $E_+$ follows Eq. (2) with no adjustable constant (slope through the origin 0.90, uncentered $R^2$ = 0.998; Sec. S11). At $G_0$ the pair's $D$ moves by at most 33 MHz (Table S6).

**The spin-imbalance relation.** The full $|E|$ is 2.12 times the hole-pair value ($R^2$ = 0.997, axes within 2.1°): the remaining orbitals, polarized by the holes, add about as much again (Sec. S11). With $m_j$ the PAW-projected spin moment on nitrogen $j$, define the spin imbalance $S_m = \sum_j m_j \exp(2\mathrm{i}\beta_j)$, with the ideal azimuths in every structure; it vanishes for any equal distribution. Then

$$E_+ = -c\, S_m\,, \tag{3}$$

with one real constant $c$ (Fig. 4). The evidence for this relation is that physically distinct perturbations share the constant (Fig. 4(c)): $c$ is 7.80–8.14 GHz per $\mu_\mathrm{B}$ for PBE donor tensor sets, 8.31 for fields and field gradients applied to the bare vacancy and 8.92 in HSE06 statics. Only the strained vacancy (10.2) and the two geometry-only tensors (6.3) depart from it (Sec. S11). Symmetry makes $E_+$ and $S_m$ proportional for any one perturbation, but not with one constant across two donor elements, relaxed lattices and applied fields, so the shared constant suggests one internal coordinate, the spin imbalance. The 36 PBE tensors, frozen and relaxed, in 8 × 8, 10 × 10 and 12 × 12 cells, are not independent; together they give $c$ = 7.95 GHz per $\mu_\mathrm{B}$ (free-phase fit: $R^2$ = 0.9987, uncentered as throughout, rms 25 MHz), across $|E|$ from 239 to 1439 MHz.

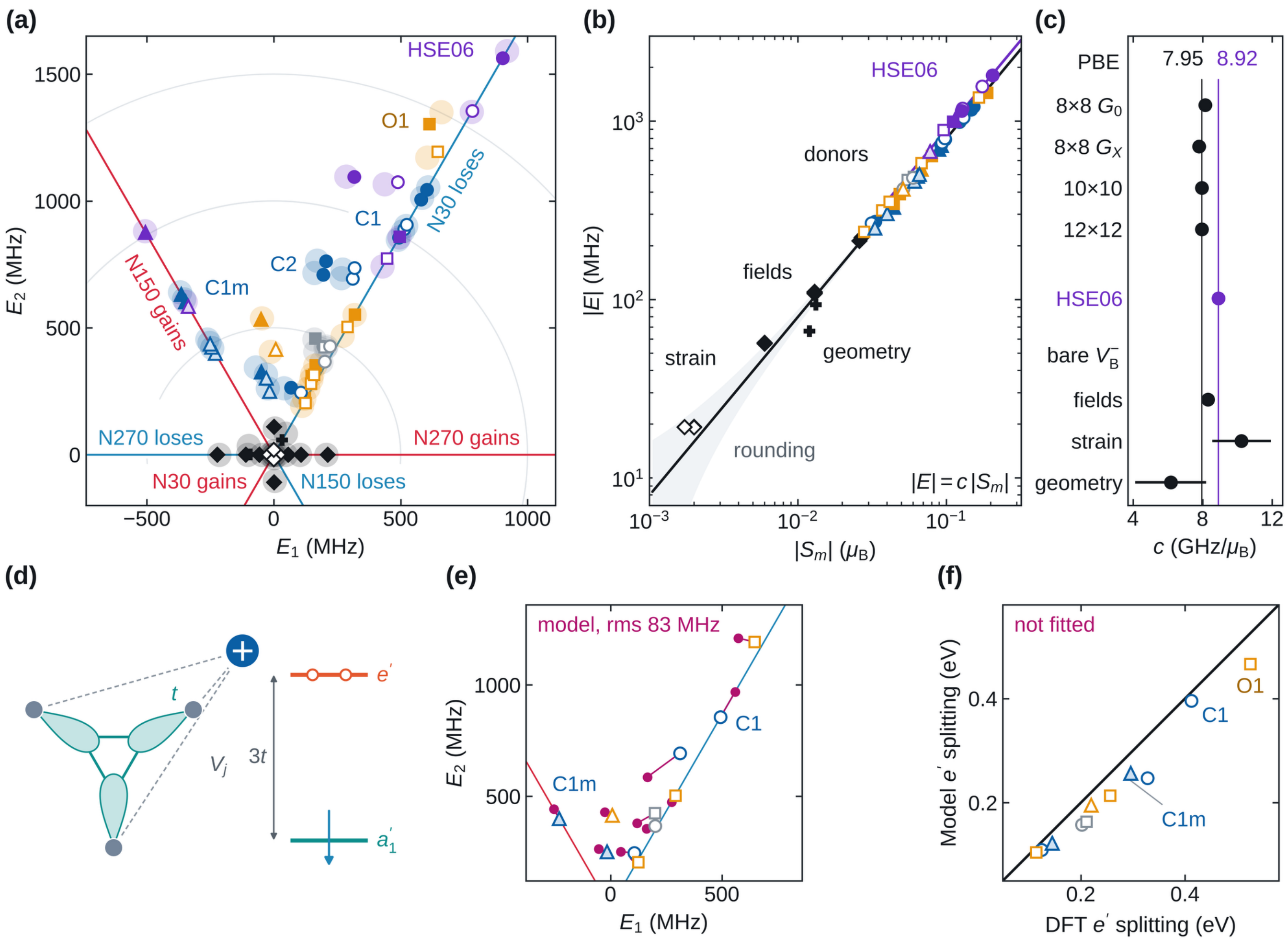

**Figure 4.** One relation for the transverse tensor. (a) $\boldsymbol{E}_+$ of every tensor (symbols as in Fig. 1; HSE06 violet; bare vacancy under strain, geometry-only distortions and applied fields including gradients, black) and $-c\boldsymbol{S}_m$ from the spin on the three nitrogens (halos), with $\boldsymbol{c}$ = 7.95 GHz $\mu_B^{-1}$ (8.92 for HSE06); the straight arms are the mirror lines. (b) $|\boldsymbol{E}|$ against $|\boldsymbol{S}_m|$; lines, $|\boldsymbol{E}| = \boldsymbol{c}|\boldsymbol{S}_m|$ with the PBE and (violet) HSE06 constants; gray band, the rounding of printed moments. (c) $\boldsymbol{c}$ per tensor set with two-standard-error intervals. (d) The three-site model of Eq. (4): potential $\boldsymbol{V}_j$ at the three nitrogens, hopping $\boldsymbol{t}$, and levels $\boldsymbol{a}_1{}'$ and $\boldsymbol{e}'$ split by $3\boldsymbol{t}$. (e) Frozen-lattice $\boldsymbol{E}$ vectors of the model ($\boldsymbol{\varepsilon}$ = 3.6, magenta, 83 MHz rms) against DFT. (f) The $\boldsymbol{e}'$ splitting of the model against DFT, not fitted.

Two cautions apply. Both sides of Eq. (3) are outputs of one calculation, so the relation describes how the tensor is carried, not how a donor position sets it. And $c$ belongs to this partition of the moments and to the ideal azimuths: with the relaxed bond azimuths, which differ by up to 1.4°, the 36 tensors give $c$ = 7.28 GHz per $\mu_B$ with 83 MHz rms.

The relation also gives the axis: when one nitrogen loses spin the $E$ axis points at it (C1, O2), and when one gains spin the axis lies across it (C1m). Nitrogen hyperfine shifts record these moves only in part (Sec. S7).

**What moves the spin.** In the three-site basis $|j\rangle$ (Fig. 4(d)), one minority electron occupies the lowest level of

$$H_{jk} = -t\left(1 - \delta_{jk}\right) + V_j\,\delta_{jk}\,, \tag{4}$$

with $t = \Delta/3 = 0.772$ eV and $V_j$ (mean $\bar{V}$) the electron's potential energy from the charge at nitrogen $j$. To first order the potential mixes $a_1{}'$ into $e'$ and shifts hole density by $\delta n_j = 2\left(V_j - \bar{V}\right)/(3\Delta)$, and within $e'$ it splits the doublet by $\delta_{e'} = \frac{2}{3}\left|\sum_j V_j \exp\left(2\mathrm{i}\beta_j\right)\right|$. Hence

$$E_+ \approx -\frac{2c_n}{3\Delta}\sum_j V_j \exp\left(2\mathrm{i}\beta_j\right), \qquad |E| \approx \frac{c_n}{\Delta}\,\delta_{e'}\,, \tag{5}$$

where $c_n$ = 5.27 GHz per hole is the constant of Eq. (3) written for hole densities, $E_+ = -c_n \sum_j \delta\, n_j \exp(2i\beta_j)$. To first order, then, $E_+$ is the second angular harmonic of the charge's external potential at the three nitrogens. Equation (5) holds on average: the DFT gap gives $c_n/\Delta$ = 2.28 GHz per eV, and the slope of $|E|$ against the DFT $e'$ splitting over the eleven frozen-lattice tensors is 2.28 ± 0.10 GHz per eV ($r$ = 0.975). Relaxation breaks this proportion: every splitting falls but, C2 apart, $|E|$ rises.

**The screening constant.** We take $t$ from the DFT gap, $c_n$ from Eq. (3) and $V_j = -e^2/(4\pi\varepsilon_0\varepsilon r_j)$ with one effective screening $\varepsilon$. Fitted to the six frozen-lattice calibration tensors, $\varepsilon$ = 3.7, and the three-site model reproduces the other five with an $E$-vector error of 56 MHz rms. Fitted to all eleven, $\varepsilon$ = 3.6, with 83 MHz rms and C3m/C3 = 0.688, against 0.600 in DFT (Fig. 4(e)). The unfitted $e'$ splittings lie 4–25% below the DFT values (Fig. 4(f)), and the bare vacancy's DFT response to a long-wavelength applied field implies a comparable $\varepsilon \approx 3.2$. The screening is fixed only together with $c_n$, and $\Delta$ and $c_n$ are effective PBE parameters (Sec. S11).

**Why direction matters.** The picture explains the frozen-lattice contrast at fixed distance (Fig. 5). C1 faces N30 at 2.82 Å, with the other two at 5.28 Å: one site near, two far. C1m faces the gap between N30 and N270, both at 3.82 Å, with N150 at 5.88 Å: two near, one far. Sampling $1/r$ at these distances, with no other input, gives a ratio of 0.554 and opposite signs of the harmonic, so the axis switches branch. The nonlinearity of Eq. (4) strengthens the contrast, because the filled orbital can localize on one near nitrogen but must share between two. The exact solution gives 0.456 (0.458 with the calibration-set screening), against 0.467 in DFT.

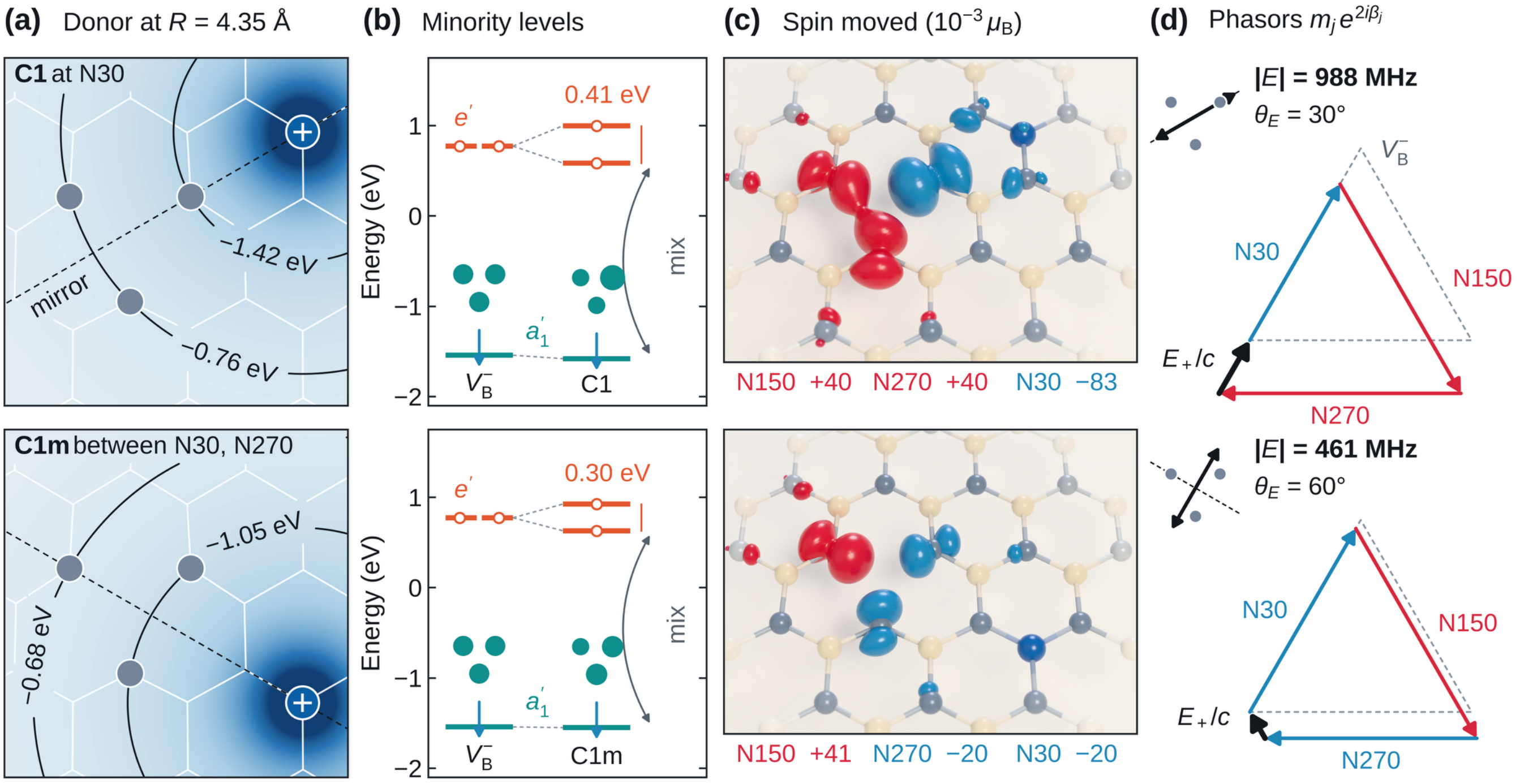

**Figure 5.** Why direction matters at equal distance: C1 against C1m (PBE, $\mathbf{8\times8}$, frozen lattice). (a) Electron potential energy of a charge $+\boldsymbol{e}$ at 4.35 Å ($\boldsymbol{\varepsilon = 3.6}$) and its values at the three nitrogens: C1 faces N30 (one near, two far), C1m the gap between N30 and N270 (two near, one far). (b) Minority levels at $\boldsymbol{\Gamma}$ (DFT; zero at the mean of the three levels): the donor splits the hole doublet $\boldsymbol{e}'$ and mixes it with $\boldsymbol{a_1}'$ (dots, its weight on the three nitrogens). (c) Spin moved, $\boldsymbol{\Delta m}$ at $\boldsymbol{\pm 0.008\,\mu_B\,\text{Å}^{-3}}$ (crimson gain, cerulean loss), and the PAW-projected change on each nitrogen. (d) The phasors $\boldsymbol{m_j \exp(2i\beta_j)}$ head to tail (dashed: the bare vacancy, where they close); the closing arrow is $\boldsymbol{E_+/c}$, and the $\boldsymbol{E}$ axis (insets) lies along the mirror for C1 (988 MHz) and across it for C1m (461 MHz).

**Why $D$ behaves differently.** Since redistribution cannot change $D$ (Eq. (2)), the $\Delta D$ of up to −80.5 MHz at $G_0$ must arise elsewhere; the hole-pair $D$ lacks the carbon–oxygen contrast there. In the geometry-only calculations, bond lengths change $D$ directly while the spin imbalance stays small: the distortion gives most of the relaxed $\Delta D$ but little of $|E|$.

## 6. Predicting and using the response

**The field-plus-gradient model.** Expanded about the vacancy center, Eq. (5) becomes a field term plus a field-gradient term. A model of this form, with fitted weights, gave the predictions we recorded before computing the controls. It treats a charged donor, chiefly a source of electric field [11], as a charge $+e$ with periodic images in a polarizable sheet with Rytova–Keldysh potential $\varphi$ and screening length $r_0$ [32]. Its field at the vacancy is $F_+ = F_x + \mathrm{i}F_y$ and its gradient $G_+ = \left(\partial_x^2\varphi - \partial_y^2\varphi\right)/2 + \mathrm{i}\ \partial_x\,\partial_y\varphi$, a rank-two term rotating like $E_+$. The $D_{3h}$ symmetry admits $F_+$ in $E_+$ only as $\mathrm{i}\kappa\bar{F}_+$ with real $\kappa$ and forbids a term in $D$ linear in $F_+$. The screened field-plus-gradient model keeps the lowest terms,

$$E_+ = \mathrm{i}\kappa\bar{F}_+ + g_{\mathrm{G}}\,G_+\,\left[+\,s_{\mathrm{geo}}\,e_{+,\mathrm{tri}}\right], \qquad \Delta D = D_0\,\left[+\,b\,\varepsilon_{\mathrm{b,tri}}\right], \tag{6}$$

where the bracketed terms, in the nitrogen triangle's deviatoric and breathing strains, act only after relaxation and $D_0$ is one constant for every donor. We fit $\kappa$, $g_{\mathrm{G}}$ and $D_0$ to 12 calibration tensors, about six independent constraints (Sec. S9). Field and gradient interfere in $|E_+|^2$ in proportion to $-\kappa g_{\mathrm{G}}\ \sin 3\psi$, so with the fitted signs they add at C1 and oppose at C1m.

The model anticipated the direction dependence. Its predicted frozen-lattice $E$ vectors miss by 8% of $|E|$ at C1m and 7% at C3m, closer than every comparison model at both, and its ratio is the 0.443 of Sec. 4, whose C1 is the model's fit (predicted C1m over computed C1: 0.504). For two carbon placements and one model form, this supports a rank-two term interfering with the linear field term, but does not single out the gradient (Sec. S9). The model does not describe $\Delta D$. It captures angular structure better than amplitude: its leave-one-out mean absolute error in frozen-lattice $|E|$ is 94.1 MHz, against 42.0 MHz for a distance power law, and at reserved C3 and O3 their errors are comparable. These tests compare the model with PBE tensors, not PBE with experiment.

**What a spectrum constrains.** Strain and other charges add an unknown offset $\delta E_+$, so a measured tensor does not fix a donor's position. If $\delta E_+$ is unconstrained, simulated spectra with 95 MHz lines [33] at signal-to-noise ratio 100 and the PBE $|E|$ scale separate no pair of configurations, under an exploratory criterion (Sec. S10). Separation comes almost entirely from an assumed 50 MHz offset distribution, a bulk-crystal $|E|$ [1]. Even then, spectra discriminate configurations, not chemistry, and without angular-control model spectra all C1m spectra go to a donor at another distance.

**Using the results.** Equations (3)–(5) estimate the frozen-lattice effect of point charges in a free-standing monolayer at the PBE $|E|$ scale, to 83 MHz rms over the eleven placements. Add their potential energies $V_j$ at the three nitrogens, solve Eq. (4) for the hole densities $n_j = 1 - a_j^2$, and read $E_+ = -c_n \sum_j \delta\, n_j \exp\left(2\mathrm{i}\beta_j\right)$ using Table II. Relaxation, which raised $|E|$ by up to 50.8%, is not included. To first order, with $\varepsilon = 3.6$, $E_+ \approx \left(6.1\ \mathrm{GHz\,Å}\right)\sum_j \exp\left(2\mathrm{i}\beta_j\right)/r_j$ for a charge $+e$ at distance $r_j$ from nitrogen $j$.

Contours in the resulting single-charge map (Fig. 6(a)) are threefold lobes: 500 MHz is reached 6.04 Å from the vacancy along a nitrogen direction but only 4.41 Å between nitrogens. Far away, $|E| \approx \mathcal{A}/R^2$ with $\mathcal{A}$ = 13.9 GHz Å$^2$ and $\theta_E \approx 45° - \psi/2$. The next order gives $|E| \approx (\mathcal{A}/R^2)[1 + (L/R)\ \sin 3\psi]$ with $L = 3r_{\mathrm{N}}/4$ = 1.15 Å, which misses $|E|$ of the eleven frozen-lattice tensors by 115 MHz rms; Eq. (5) misses it by 84 MHz and $\theta_E$ by 4.1°.

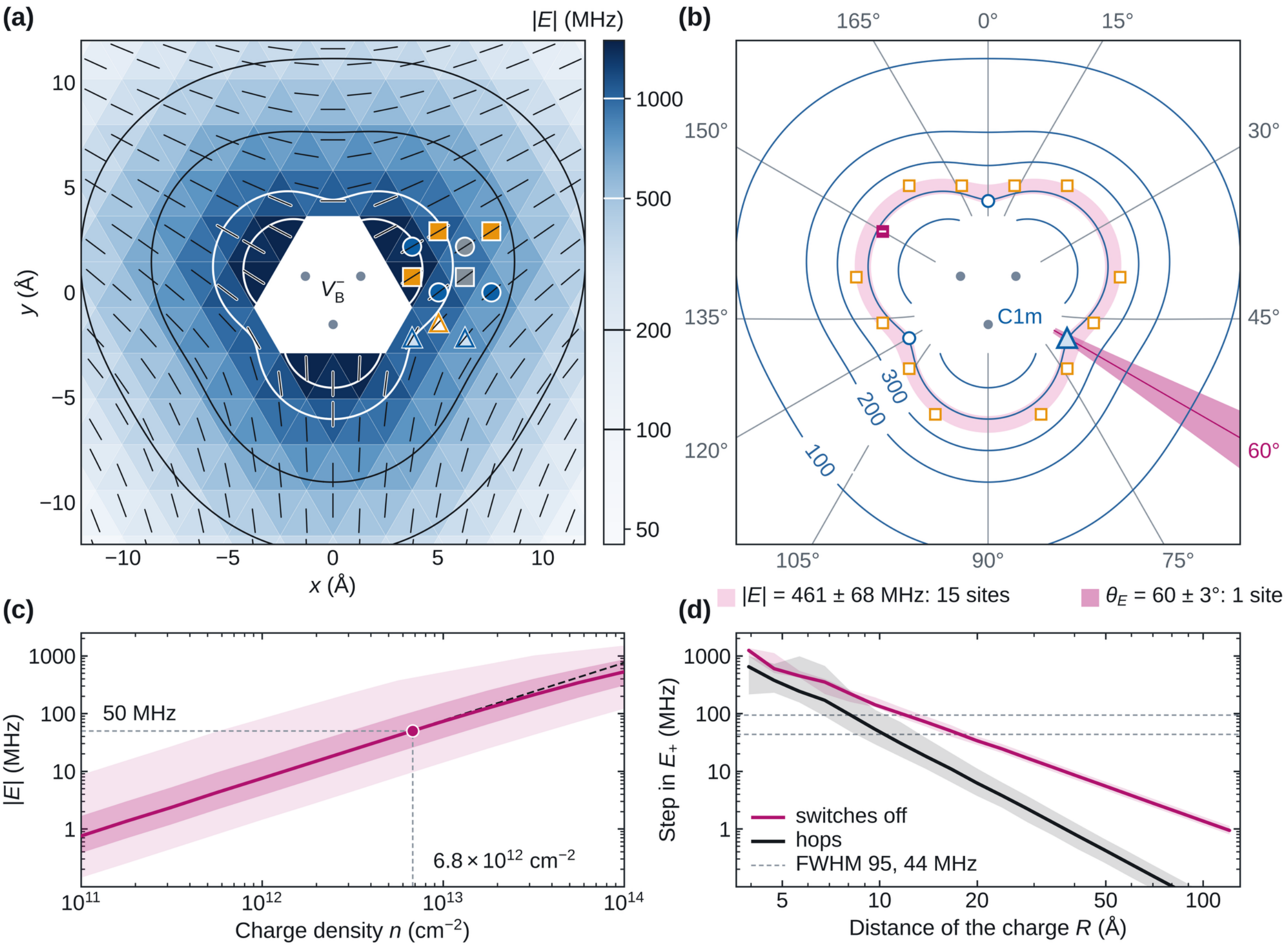

**Figure 6.** Using the results. Three-site model, Eq. (4), with $\varepsilon$ = 3.6, for a free-standing monolayer at the PBE $|E|$ scale: an illustration, not a prediction for a measured sample. (a) $|\boldsymbol{E}|$ for one charge $+\boldsymbol{e}$ on each lattice site (color and contours, logarithmic) and $\boldsymbol{E}$ axis (ticks); symbols, placements (Fig. 1(b)). (b) A measurement inverted on iso-$|\boldsymbol{E}|$ lines (MHz) and iso-$\boldsymbol{\theta_E}$ rays: for $|\boldsymbol{E}|$ = 461 ± 68 MHz, 15 sites qualify (band), and $\boldsymbol{\theta_E}$ = 60 ± 3° leaves one (wedge); filled square, a negative charge giving the same reading. (c) Ensemble $|\boldsymbol{E}|$ against the density $\boldsymbol{n}$ of random charges: median, quartiles, 5–95% range; dashed, dilute limit. (d) Step in $\boldsymbol{E_+}$ when one charge at distance $\boldsymbol{R}$ switches off (magenta) or hops one lattice constant (black) amid charges at the 50 MHz density (medians; bands, 5–95%); dashed, 95 and 44 MHz linewidths.

Used in reverse (Fig. 6(b)), $|E|$ alone selects a threefold curve of sites, and the axis cuts it to a narrow wedge, opposite for a negative charge. The axis should therefore always be reported; measuring it needs an in-plane field or a single defect aligned against the lattice. For C1m (461 MHz), 15 sites lie within the three-site model's 68 MHz rms error in $|E|$, and without an offset its 3.2° rms axis error leaves one. This case is favorable: C1m entered the $\varepsilon$ fit, and only 7 of 11 placements are recovered at their computed site (Sec. S11).

For random charges of both signs at density $n$, the median ensemble $|E|$ is $\sqrt{3}\pi n\mathcal{A}$ at low density, 7.6 MHz per $10^{12}$ cm$^{-2}$. Read as this median, the 50 MHz bulk-crystal $|E|$ corresponds to $n = 6.8 \times 10^{12}$ cm$^{-2}$ in this model (Fig. 6(c)), or $2.3 \times 10^{12}$–$1.4 \times 10^{13}$ cm$^{-2}$ across our screening choices (Sec. S11). The model likely overestimates these densities, and strain also contributes in real samples. At that density a charge switching off within 16.6 Å moves $E_+$ by more than 50 MHz in the median, and one hopping by a lattice constant does so within 9.9 Å (Fig. 6(d)).

For a new defect, the spin on the three sites is worth reporting with its partition; here the dimensionless $c\,m_0/D$ is 1.26 (Table II; 1.67 in HSE06 statics), untested on another defect. Ratios at fixed distance are safer than absolute $|E|$, and $\Delta D$ needs relaxed structures.

**Table II.** Constants of the spin response of $V_B^-$ in monolayer hBN, for use with Eqs. (2)–(6). Uncorrected PBE, $8 \times 8$ cell, unless noted; $b$ and $s$ are clamped-ion. The moment is the PAW-projected value (Sec. S4) and $\beta_j$ takes its ideal values, so $c$ belongs to that definition; $c_n$ refers to hole weights on the three nitrogens, scaled by the one factor giving two holes in the bare vacancy. The PBE $c$ pools the 36 tensors of all cell sizes (Sec. 5); its uncertainty is a bootstrap interval over tensors, not a method error. Closed-form couplings follow from Eq. (5) per unit local field, the fitted ones per unit screened model field, the bare ones per unit DFT-screened field (Secs. S9 and S11).

| Quantity | Symbol | PBE | Other | Unit |
|---|---|---|---|---|
| Axial zero-field parameter | $D$ | 2798.8 | 2551.4 (HSE06), 3480 (expt.) | MHz |
| Spin moment per nitrogen | $m_0$ | 0.445 | 0.477 (HSE06) | $\mu_B$ |
| Minority $a_1'$–$e'$ gap | $\Delta$ | 2.315 | 3.31 (HSE06) | eV |
| Three-site hopping, $\Delta/3$ | $t$ | 0.772 | | eV |
| Spin-imbalance constant | $c$ | 7.95 (7.85–8.05) | 8.92 (HSE06), 8.31 (fields, gradients) | GHz $\mu_B^{-1}$ |
| Hole-imbalance constant | $c_n$ | 5.27 | 5.41 (calibration only) | GHz per hole |
| $\|E\|$ per $e'$ splitting | $c_n/\Delta$ | 2.28 | 2.28 ± 0.10 (fit, $G_0$) | GHz eV$^{-1}$ |
| Effective screening, three-site | $\varepsilon$ | 3.6 | 3.7 (calibration only) | |
| Field coupling | $\kappa$ | −2.96 | −3.08 (bare), −3.48 (closed form) | GHz (V Å$^{-1}$)$^{-1}$ |
| Gradient coupling | $g_G$ | 3.81 | 3.12 (bare), 2.66 (closed form) | GHz (V Å$^{-2}$)$^{-1}$ |
| Biaxial strain coupling of $D$ | $b$ | −12.06 | | GHz per unit strain |
| Shear strain coupling of $E$ | $s$ | 19.2 | | GHz per unit strain |
| Empirical relaxation term | $s_{geo}$ | −8.155 | | GHz per unit strain |
| Sheet screening length | $r_0$ | 7.08 | 10.09 (relaxed ions) | Å |

## 7. Conclusions

At fixed distance from the boron vacancy, a carbon donor's direction alone can halve the frozen-lattice $|E|$ and turn the $E$ axis across a mirror line. Tests fixed in advance support this direction dependence; its explanation is exploratory. Two holes in the $e'$ doublet carry $D$ and about half of $|E|$, and the orbitals they polarize add the rest. A charge moves spin around the nitrogen triangle, and $E_+$ follows this imbalance through one constant. In PBE, redistribution leaves $D$ almost untouched, so $E$ reflects a charge's position, whereas $\Delta D$ combines bond-length changes with a frozen-lattice shift the picture does not explain. A three-site model with one screening constant maps a charge's position onto the frozen-lattice $E_+$.

Two predictions are testable for single vacancies near a carbon donor of known position, if the optical readout keeps the computed charge configuration: at equal vacancy-center distance $|E|$ is lower where $\sin 3\psi$ is negative, and on a mirror line the $E$ axis switches branch. Our periodic, static calculations treat one charge state in a free-standing monolayer. The next steps are a second charge state, with capture and emission rates, and screened calculations for shallow vacancies in thicker flakes [34]. Deriving Eq. (2) needs a doublet shared by three equivalent sites, the level scheme of nitrogen-vacancy-like spin defects [35,36], so the picture may extend to other defects with a comparable three-site doublet; we have not tested it there.

## Acknowledgments

This work used the Anvil supercomputer at Purdue University through allocation MAT260089 from the Advanced Cyberinfrastructure Coordination Ecosystem: Services & Support (ACCESS) program, which is supported by U.S. National Science Foundation grants 2138259, 2138286, 2138307, 2137603, and 2138296 [37,38]. This research also used resources of the Oak Ridge Leadership Computing Facility at Oak Ridge National Laboratory, which is supported by the Office of Science of the U.S. Department of Energy under Contract No. DE-AC05-00OR22725.

## Data availability

The data that support the findings of this article are openly available on Zenodo under DOI 10.5281/zenodo.23575852 [39]. The deposit holds the inputs, structures and output files of every production calculation and the data behind every table and figure (Sec. S2). The licensed PAW potential files and our two patches to the licensed VASP source are not redistributed; the deposit identifies each by name and checksum.

# Supplemental Material

*Theory of the charge response of the boron-vacancy spin in monolayer hexagonal boron nitride*

Dongyeon (Tyler) Kim[1], Theodore Scharlotta[2], and Daniel Hashemi[1*]

[1] Department of Physics, Optical Engineering, and NanoEngineering, Rose-Hulman Institute of Technology, 5500 Wabash Avenue, Terre Haute, Indiana 47803, USA
[2] Department of Chemical Engineering, Rose-Hulman Institute of Technology, 5500 Wabash Avenue, Terre Haute, Indiana 47803, USA
*Contact author: hashemi@rose-hulman.edu
ORCID: Dongyeon (Tyler) Kim 0009-0000-4616-2695; Theodore Scharlotta 0009-0007-7485-717X; Daniel Hashemi 0000-0001-7328-1032

## S1. Cells, placements and conventions

This Supplemental Material provides the computational details and additional tests supporting the main results. The data deposit on Zenodo, DOI 10.5281/zenodo.23575852 [39], holds the raw data and output files.

Cells. All defect cells are supercells of one PBE primitive cell with $\mathbf{a}_1 = a(1,0,0)$ and $\mathbf{a}_2 = a(1/2, \sqrt{3}/2, 0)$ (B at (0, 0), N at (1/3, 1/3)), $a$ = 2.5130 Å and a 25 Å cell height. The 8 × 8, 10 × 10 and 12 × 12 supercells hold 127, 199 and 287 atoms, donor or not. $V_{\mathrm{B}}^{-}$ carries $q = -1$. A vacancy–donor complex is neutral, because the ionized donor supplies the vacancy's extra electron, and it keeps the 510, 798 or 1150 electrons of $V_{\mathrm{B}}^{-}$.

Placements. $R$ and $\psi$ are the donor's in-plane minimum-image distance and azimuth, counterclockwise from $\mathbf{a}_1$, measured from the centroid of the three vacancy nitrogens (Table S1); the paper uses $R(G_0)$. Relaxation shifts them by at most 0.026 Å and 0.03° in the 8 × 8 cell, and no geometry moves a vacancy nitrogen more than 1.4° from its named azimuth. At $G_0$, N30 is the vacancy nitrogen nearest the donor for C1–C4 and O1–O4 (2.46–6.58 Å) and for C3m and Om. C1 and O2 lie on the mirror line through N30; C1m lies on the mirror line through N150, where N30 and N270 tie (3.82 Å).

Distinct placements. The $D_{3h}$ operations of $V_{\mathrm{B}}^{-}$ (mirror lines at 30°, 90° and 150°) map each site onto an orbit of three or six sites, one of which lies in $-30° \le \psi \le 30°$. C1m and C3m have the ideal distances of C1 and C3 ($\sqrt{3}\,a$ = 4.3527 Å, $\sqrt{7}\,a$ = 6.6488 Å) but belong to other orbits, because $\psi \to -\psi$ is not a symmetry of the defect. Every nitrogen shell within 9 Å is a single orbit, so no $O_{\mathrm{N}}$ site, including the control Om (5.24 Å, $\sin\, 3\psi = -0.75$), has such a partner.

Spin Hamiltonian. Each configuration is described by

$$\mathcal{H}/h = \mathbf{S}\cdot\mathbf{D}\cdot\mathbf{S} + \gamma_{\mathrm{e}}\,\mathbf{B}\cdot\mathbf{S} + \sum_k \left(\mathbf{S}\cdot\mathbf{A}_k\cdot\mathbf{I}_k + \mathbf{I}_k\cdot\mathbf{Q}_k\cdot\mathbf{I}_k - \gamma_{\mathrm{N}}\,\mathbf{B}\cdot\mathbf{I}_k\right), \tag{S1}$$

with $\mathbf{A}_k$ and $\mathbf{Q}_k$ the hyperfine and quadrupole tensors of nucleus $k$, $\gamma_{\mathrm{e}} = g\mu_{\mathrm{B}}/h$ with isotropic $g$, and $\gamma_{\mathrm{N}}$ the $^{14}$N gyromagnetic ratio. $\mathbf{D}$ is the spin–spin term of the unrestricted Kohn–Sham determinant with $S_z = 1$, evaluated with the periodic method of Rayson and Briddon [23], without spin–orbit coupling.

Principal axes. The normal is always a principal axis ($|D_{xz}|, |D_{yz}| \le 0.001$ MHz), so for $\mathbf{B}$ along it the transitions lie at $\nu_{\pm} = \left|D \pm \sqrt{(\gamma_{\mathrm{e}}B)^2 + |E|^2}\right|$. Sorting the axes by magnitude would move $z$ into the plane for C1 ($|E|/D$ = 0.363 at $G_0$, 0.422 at $G_X$) and O1 (0.488, 0.533); we do not, which relabels parameters but not the spectrum. On a mirror line the calculated sign selects the branch: $\theta_E$ = 30° or −60° on the 30° line (C1, O2) and 60° or −30° on the −30° line (C1m).

Nuclei. Hyperfine and quadrupole tensors come from the all-electron density reconstructed by the PAW method [21,40], for $^{14}$N ($I = 1$, $\gamma_N$ = 3.077705 MHz/T) and the spinless donors $^{12}$C and $^{16}$O [41]. The primary hyperfine tensor keeps the plane-wave and one-center contact terms, the core-polarization term and the dipolar part [40]. For the quadrupole tensor $Q_{ij} = eQV_{ij}/[2I(2I-1)h]$ we use $Q(^{14}\mathrm{N})$ = 20.44 mb [42] and the computed sign of the field gradient, not checked against a reference of known sign.

**Table S1.** Where each donor sits and whether the vacancy keeps its triplet (PBE, $\boldsymbol{S_z = 1}$ fixed; paired entries are $\boldsymbol{G_0}$ / $\boldsymbol{G_X}$), with the relaxed angular controls in the second entry. The donor sits at $\boldsymbol{n\mathbf{a}_1 + m\mathbf{a}_2}$ from the vacancy ($\boldsymbol{C_B}$) or at $(\boldsymbol{n+1/3})\,\mathbf{a}_1 + (\boldsymbol{m+1/3})\,\mathbf{a}_2$ ($\boldsymbol{O_N}$); $\boldsymbol{R}$ and $\boldsymbol{\psi}$ from the vacancy-nitrogen centroid. $\boldsymbol{f_{3N}}$: share of the PAW-projected moment on the three vacancy nitrogens; defect gap: highest occupied to lowest empty level, set at $\boldsymbol{G_0}$ by an empty carbon level below the vacancy pair for C4, C3 and C3m. All tensors pass the state check (Sec. S3).

| Site | Sublattice | $(n, m)$ | $R$ (Å) | $\psi$ (°) | $f_{3N}$ | Donor moment ($\mu_B$) | Defect gap (eV) |
|---|---|---|---|---|---|---|---|
| *Reference (no donor)* | | | | | | | |
| $V_B^-$ $(8 \times 8)$ | – | – | – | – | 0.849 | – | 1.65 |
| *Calibration placements (fit the model of Sec. 6)* | | | | | | | |
| C1 | B | (1, 1) | 4.349 / 4.360 | 30.00 / 30.00 | 0.851 / 0.847 | −0.006 / −0.004 | 1.33 / 1.46 |
| C2 | B | (2, 0) | 5.023 / 5.023 | 0.02 / 0.06 | 0.850 / 0.851 | 0.001 / 0.001 | 1.40 / 1.45 |
| C4 | B | (3, 0) | 7.538 / 7.537 | 0.01 / 0.02 | 0.854 / 0.852 | 0.000 / 0.000 | 0.97 / 1.56 |
| O1 | N | (1, 0) | 3.850 / 3.876 | 10.90 / 10.93 | 0.844 / 0.845 | 0.012 / 0.017 | 1.17 / 1.29 |
| O2 | N | (1, 1) | 5.810 / 5.820 | 30.00 / 30.00 | 0.854 / 0.851 | 0.001 / 0.001 | 1.49 / 1.56 |
| O4 | N | (2, 1) | 8.082 / 8.086 | 21.06 / 21.06 | 0.853 / 0.852 | 0.000 / 0.000 | 1.58 / 1.61 |
| *Reserved: computed after the predictions were recorded (relaxed structures inspected before)* | | | | | | | |
| C3 | B | (2, 1) | 6.648 / 6.652 | 19.12 / 19.13 | 0.854 / 0.852 | −0.002 / −0.001 | 1.14 / 1.57 |
| O3 | N | (2, 0) | 6.330 / 6.337 | 6.60 / 6.62 | 0.853 / 0.854 | 0.001 / 0.001 | 1.51 / 1.56 |
| *Angular controls: reversed azimuth (C1m and C3m at the distances of C1 and C3; Om between the O1 and O2 distances)* | | | | | | | |
| C1m | B | (2, −1) | 4.347 / 4.343 | −30.00 / −30.00 | 0.849 / 0.853 | −0.005 / −0.003 | 1.34 / 1.53 |
| C3m | B | (3, −1) | 6.646 / 6.642 | −19.10 / −19.09 | 0.853 / 0.851 | −0.001 / 0.000 | 1.16 / 1.58 |
| Om | N | (2, −1) | 5.237 / 5.247 | −16.09 / −16.08 | 0.852 / 0.851 | 0.000 / 0.000 | 1.50 / 1.56 |
| *Larger $10 \times 10$ cell (finite-size check)* | | | | | | | |
| $V_B^-$ $(10 \times 10)$ | – | – | – | – | 0.849 | – | 1.65 |
| C2 $(10 \times 10)$ | B | (2, 0) | 5.023 / 5.022 | 0.03 / 0.07 | 0.849 / 0.851 | 0.001 / 0.001 | 1.36 / 1.44 |
| O4 $(10 \times 10)$ | N | (2, 1) | 8.084 / 8.089 | 21.06 / 21.06 | 0.854 / 0.853 | 0.000 / 0.000 | 1.55 / 1.60 |

## S2. Computational settings and data

Settings. We use VASP 6.6.1 [20,21] with the PBE functional [19] and the code's standard PBE PAW datasets, with valences 3, 5, 4 and 6 for B, N, C and O. The plane-wave cutoff is 550 eV, with nonspherical gradient corrections inside the atomic spheres, no symmetry imposed and Gaussian smearing of 0.02 eV. Supercells are spin polarized, sample Γ only and truncate the Coulomb interaction along the sheet normal [22], so in-plane images are what remains of the finite-size problem. The lattice parameter is the minimum of a quadratic through primitive-cell energies on a $21 \times 21 \times 1$ mesh.

Relaxations keep the cell fixed and stop when the energy is self-consistent to $10^{-7}$ eV and every force is below 0.001 eV/Å; static and tensor calculations converge to $10^{-8}$ eV. Every production calculation starts from 2/3 $\mu_B$ on each vacancy nitrogen with the total spin fixed at $S_z = 1$, which selects the high-spin sector but does not establish a ground state. All tensor calculations use Broyden mixing.

Checks. Every production calculation passed automated checks: verified inputs, normal termination, a converged final electronic loop and complete output. An independently written reader reproduces $D$ and $|E|$ of four representative calculations within 0.05 MHz. The electrostatic model was re-implemented independently, and the analysis code could read the reserved calculations only after their predictions had been recorded (Sec. S9).

Data. The deposit, DOI 10.5281/zenodo.23575852, holds the inputs, structures and tensors of every production calculation, with check and metadata files where they were written, the VASP output files with the licensed potential text removed, the archived predictions and the data behind every table and figure, with PAW dataset identities in place of the licensed potential files. It complements open databases of two-dimensional defects [43,44], our study of vacancy pairs in monolayer hBN [45] and our dataset on the neutral carbon-dimer triplet in strained monolayer hBN [46] with $V_B^-$ tensors resolved by donor chemistry and placement. The analysis used NumPy [47], SciPy [48], pymatgen [49] and pawpyseed [50], and the figures used Matplotlib [51], scikit-image [52] and Blender.

## S3. Electronic-state checks

Every reported PBE tensor passes the checks for the vacancy's two-hole triplet. A tensor counts as the vacancy $S = 1$ state only if it passes five criteria, the state check, on the Γ-point levels and PAW-projected moments (Sec. S4): (1) exactly two empty minority-spin and no empty majority-spin vacancy levels, each with more than 0.8 of its vacancy-nitrogen s and p weight in in-plane p orbitals; (2) $f_{3N} \geq 0.8$, the share of the moment on the three vacancy nitrogens; (3) in a complex, a donor moment below 0.05 $\mu_B$ in magnitude and no donor-localized level occupied by 0.5 or more; (4) integer occupations and a defect gap above 0.5 eV; (5) for a relaxed geometry, a maximum force of at most 0.001 eV/Å.

All checked PBE tensors pass (Table S1, Fig. S1(a)), as do the angular controls and the 8 larger-cell tensors of the angular pairs. In PBE the lowest empty carbon level lies 0.97–1.85 eV above the highest occupied level at $G_0$ and 1.87–2.66 eV at $G_X$, near or below the 2.33 eV of 532 nm excitation. O2–O4 levels lie at 2.68–3.40 eV, and none is identified for O1, whose level drawn in Fig. 2(c) sits on its boron neighbors. These are Kohn–Sham differences, not charge-transition levels or optical thresholds, and no rate is computed.

Within the tested starts, no state lies below the triplet (Fig. S1(b)). At $V_B^-$, C2 and O2, starts with the total spin free return the triplet, and with $S_z = 0$ imposed they converge to broken-symmetry determinants above it. The closest donors, C1 and O1, had nine starts each at $G_0$ and $G_X$. All 20 that leave the spin free or fix it at $S_z = 1$ collapse onto the triplet, and the 16 held at $S_z = 0$ reach nine distinct solutions 0.155–0.758 eV above it, vertical differences, not singlet–triplet gaps. None lies within 25 meV, the margin for relaxation.

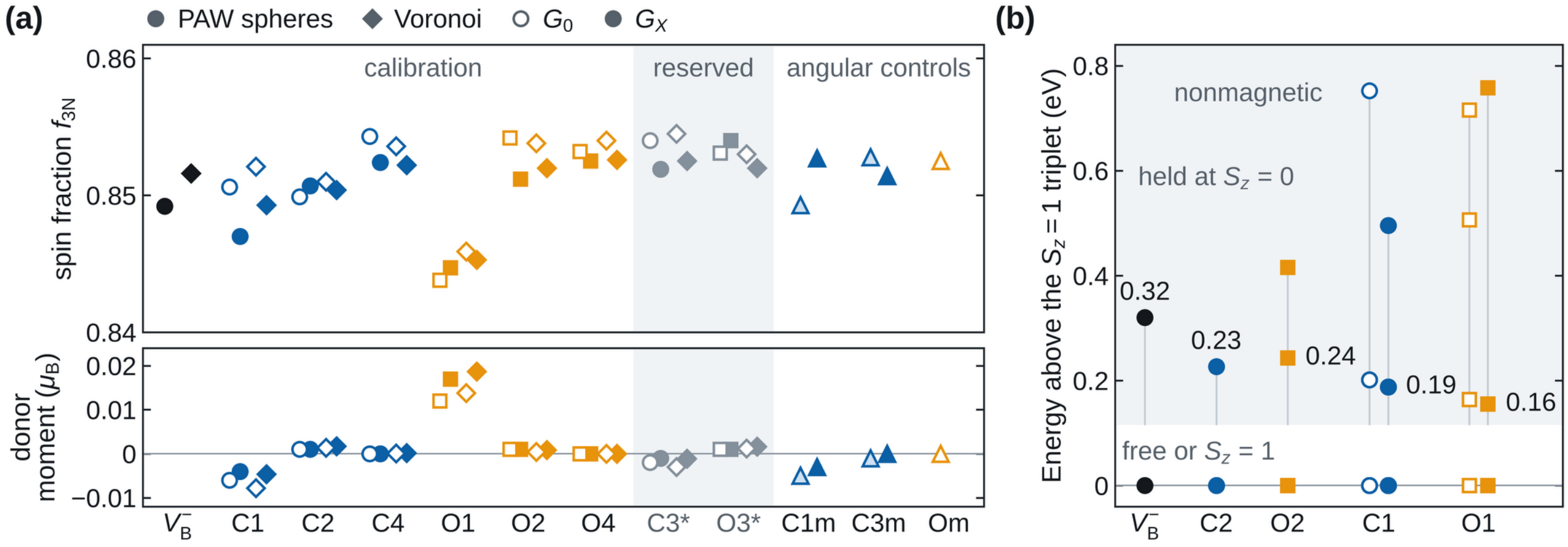

**Figure S1.** State checks (PBE, $8 \times 8$; open $G_0$, filled $G_X$; *, reserved). (a) Share $f_{3N}$ of the spin moment on the three vacancy nitrogens (top) and moment on the donor (bottom), within PAW spheres (circles, squares; triangles for the angular controls) and Voronoi cells (diamonds). (b) Restart energies above the $S_z = 1$ triplet of each complex. All runs left free or fixed at $S_z = 1$ return to the triplet (white floor). Runs held at $S_z = 0$ lie above it (tinted), with the lowest energy of each complex printed; the three near 0.75 eV are nonmagnetic. Broken-symmetry values are vertical energy differences, not singlet–triplet gaps.

## S4. Spin-density partitions

The paper uses two spin populations and never equates them. PAW-projected moments are the spin that the atom-centered projectors capture inside each atom's augmentation sphere (0.79 Å for N, set by the PAW datasets) [21]. We also integrate the magnetization density $m = n_{\mathrm{maj}} - n_{\mathrm{min}}$ (pseudo-density plus compensation charge) over spheres of the datasets' Wigner–Seitz radii (0.741 Å for N) and over Voronoi cells (Fig. S2). The measures agree on where the spin sits. Sphere and PAW-projected moments of a vacancy nitrogen differ by a nearly constant factor (0.897–0.903), and the three vacancy-nitrogen Voronoi cells, which include interstitial spin, hold 0.845–0.854 of the full moment, against a PAW-projected share of 0.844–0.854.

The donor moves spin among the vacancy nitrogens but carries none itself: the donor atom holds −0.008 to 0.019 $\mu_{\mathrm{B}}$ in its Voronoi cell. At $G_0$ the N30 sphere loses 0.016–0.104 $\mu_{\mathrm{B}}$ and N150 and N270 gain 0.007–0.048 $\mu_{\mathrm{B}}$, while the three Voronoi cells together change by only −0.012 to +0.005 $\mu_{\mathrm{B}}$. The PAW-projected changes are 1.12 times the sphere changes (correlation 1.000, 48 values). These ranges cover calibration placements only.

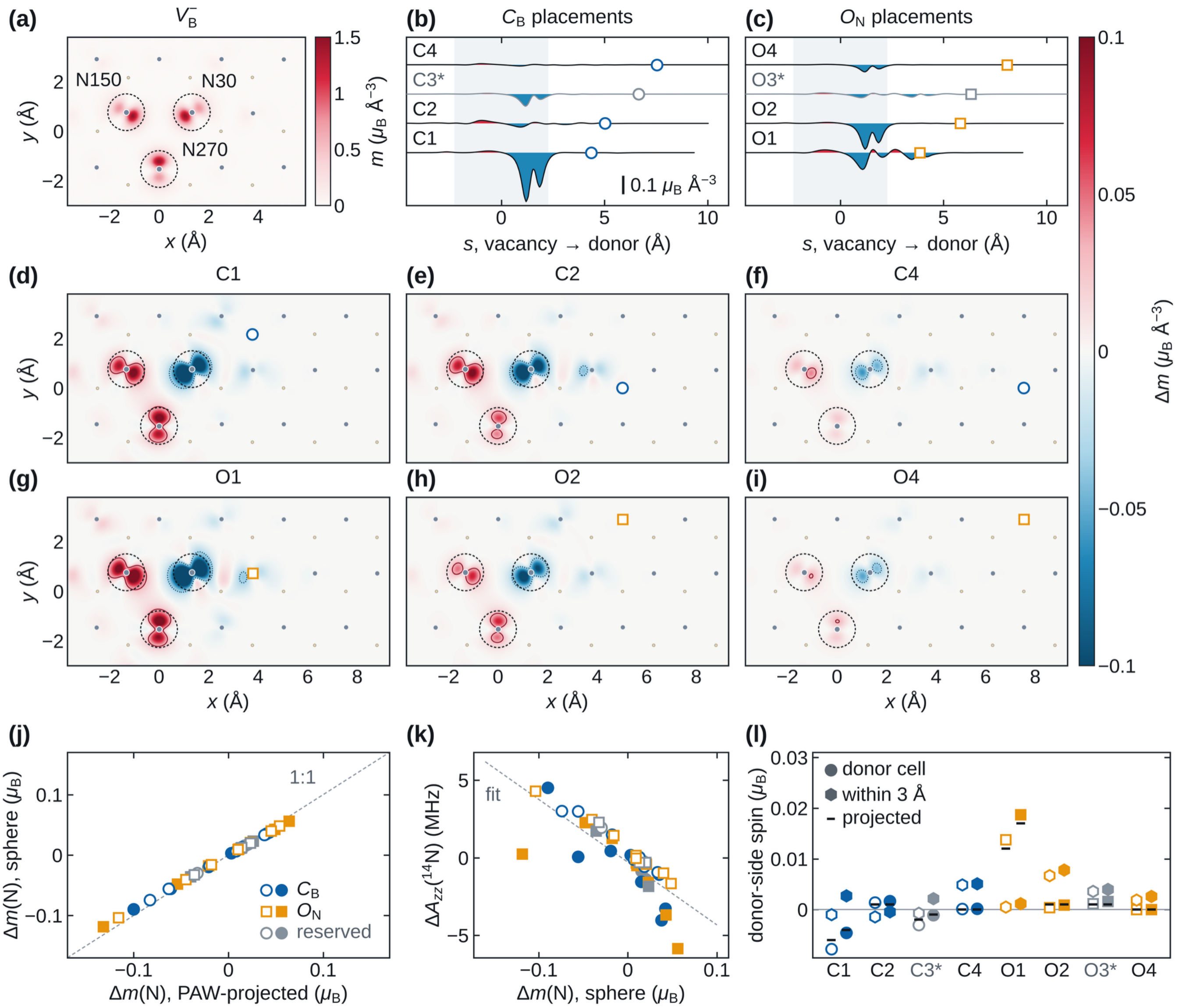

**Figure S2.** Spin density of $V_B^-$ and its change next to a donor (PBE, $8\times8$, $\Gamma$ point). Maps: magnetization density in the atomic plane, $x \parallel \mathbf{a}_1$; dashed circles: 0.741 Å integration spheres. (a) $m(\mathbf{r})$ of $V_B^-$; each sphere holds 0.401 $\mu_B$ (PAW-projected moment 0.445 $\mu_B$), each Voronoi cell 0.568 $\mu_B$. (b, c) $\Delta m = m_X(G_0) - m_{V_B}$ along the vacancy–donor line; tint, reach of vacancy-nitrogen spheres. (d–i) $\Delta m$ maps of the six calibration placements; contours at $\pm 0.03\,\mu_B$ Å$^{-3}$. N30 loses spin (cerulean); N150 and N270 gain it (crimson). (j) Sphere against PAW-projected change. (k) $\Delta A_{zz}$ (core included) against sphere change; dashed: calibration fit, slope −40 MHz $\mu_B^{-1}$, $r = -0.815$ ($n = 36$). (l) Donor spin: Voronoi cell, symbols; PAW-projected, bars; donor-side cells within 3 Å, hexagons. Symbols: Fig. 1.

## S5. Finite size

Cell size acts on $|E|$ far more than on $D$. From $8 \times 8$ to $10 \times 10$, $D$ of $V_B^-$ falls by 24.8 MHz and $\Delta D$ of C2 and O4 changes by +0.4 to +12.5 MHz, whereas $|E|$ grows by 5.5–7.3% for the near donor C2 and 32.0–35.3% for the far donor O4.

We distinguish three kinds of estimate. (a) Periodic tensors: $8 \times 8$ for every placement, $10 \times 10$ for C2, O4 and the angular pairs (Table S2), and $12 \times 12$ for O4 and C1/C1m. (b) Isolated-complex model estimates extrapolate the $8 \times 8$ DFT vector with the model's image term (Sec. S9), multiplicatively (the default) or additively,

$$E_{+,\infty} = E_+^{\mathrm{DFT}} \frac{E_{+,\infty}^{\mathrm{model}}}{E_+^{\mathrm{model}}} \quad \text{or} \quad E_{+,\infty} = E_+^{\mathrm{DFT}} + E_{+,\infty}^{\mathrm{model}} - E_+^{\mathrm{model}}. \tag{S2}$$

(c) The variant set, the model variants with $r_0$ between 0 and 10.09 Å that reproduce the four $10 \times 10$ $|E|$ of C2 and O4 to within 3.3%, measures the sensitivity of (b) to screening length, model form and extrapolation method. These tensors also served to choose the model, so they do not validate the isolated-complex limit.

The model's step covers 44–51% of its isolated-complex correction of $|E|$; the rest is extrapolated. The variant range of isolated $|E|$ for a calibration tensor is 34–57 MHz wide (Table S2) and cancels only partly in contrasts: the C4/O4 $G_X$ contrast spans −0.4 to +42.4 MHz, so its sign is not determined. Multiplicative estimates put the frozen-lattice ratios at 0.51 for C1m/C1 and 0.70 for C3m/C3. Post hoc, estimates with the image terms of the nonlinear variant (0.418, 0.534) and the field-only model (0.467, 0.600) also keep them below 1, but whether images strengthen or weaken the contrast depends on the model.

O4 at $12 \times 12$. Before any $12 \times 12$ calculation existed, the variant set, anchored at $10 \times 10$, predicted $|E|$ = 346.5–358.5 MHz at $G_0$ and 375.1–387.1 MHz at $G_X$, with no pass criterion. The calculations give 352.6 MHz, inside the range, and 389.4 MHz, 2.3 MHz above it; the $G_X$ prediction used the $10 \times 10$ relaxed geometry. A post hoc power-law fit through the three sizes gives isolated-complex limits of 414 MHz at $G_0$, inside its variant range, and 474 MHz at $G_X$, 19 MHz above it (Table S2), so the set is not a comprehensive bound. At $12 \times 12$ O4 has $\Delta D = -10.8$ / $-12.7$ MHz ($-12.2$ / $-14.4$ MHz at $10 \times 10$), a post hoc cell-level trend.

Angular pairs. Under criteria fixed in advance, C1 and C1m were computed at their frozen-lattice positions in larger supercells of the relaxed vacancy. Their $|E|$ is 1027.5 / 490.2 MHz at $10 \times 10$ (ratio 0.477, +2.1% from $8 \times 8$) and 1046.6 / 503.6 MHz at $12 \times 12$ (0.481, +3.0%), with $\theta_E$ = 30.00° / 60.00° in both. An $L^{-2.7}$ fit in lateral size $L$, a form fixed before the runs, gives an isolated-complex estimate of 0.488 (sensitivity range 0.484–0.493 for exponents 2–4 and a free exponent). The computed ratios lie inside the ranges recorded beforehand (0.467–0.490, 0.467–0.501). Relaxed at $10 \times 10$, the pair gives 0.606 (+1.0%; recorded five-variant range 0.600–0.617) and the C1m axis keeps its 60° branch. The frozen-lattice C3m/C3 at $10 \times 10$ is 0.630 (+5.0%; recorded range 0.600–0.650), below the qualitative 0.800 threshold fixed in advance, and both axes turn by at most 0.95°. The relaxed C3/C3m pair has no larger-cell result.

**Table S2.** Finite-size data for the zero-field splitting (PBE, MHz; paired entries are $\boldsymbol{G_0}$ / $\boldsymbol{G_X}$, angular controls at $\boldsymbol{G_0}$ except Om). The $\mathbf{10 \times 10}$ columns give the periodic tensors of the larger cell; $\boldsymbol{\Delta D}$ is relative to $\boldsymbol{V_B^-}$ of the same cell. Isolated $|\boldsymbol{E}|$: model estimates from the $\mathbf{8 \times 8}$ tensors ($\boldsymbol{r_0}$ = 7.08 Å), extrapolated by Eq. (S2); variant-set range: spread over all variants. Reserved and angular-control rows are descriptive.

| Donor | $\lvert E\rvert$, $8 \times 8$ | $\lvert E\rvert$, $10 \times 10$ | $\theta_E$, $10 \times 10$ (°) | $\Delta D$, $10 \times 10$ | Isolated $\lvert E\rvert$, multiplicative | Isolated $\lvert E\rvert$, additive | Variant-set range of isolated $\lvert E\rvert$ |
|---|---|---|---|---|---|---|---|
| $V_B^-$ | 0.006 | 0.063 | – | – | – | – | – |
| *Calibration placements (fit the model of Sec. 6)* | | | | | | | |
| C1 | 987.6 / 1160.5 | 1027.5 / 1205.8 | 30.0 / 30.0 | −78.2 / −36.5 | 1070.3 / 1248.9 | 1081.6 / 1254.8 | 1048–1092 / 1232–1265 |
| C2 | 759.9 / 735.7 | 802.0 / 789.5 | 33.3 / 37.5 | −53.7 / −106.2 | 877.6 / 850.8 | 861.2 / 842.3 | 836–890 / 809–863 |
| C4 | 267.4 / 273.2 | – | – | – | 440.1 / 451.8 | 433.7 / 445.8 | 399–453 / 410–465 |
| O1 | 1357.4 / 1439.4 | – | – | – | 1444.2 / 1525.4 | 1436.7 / 1520.6 | 1414–1455 / 1500–1536 |
| O2 | 581.0 / 637.6 | – | – | – | 695.8 / 752.7 | 708.0 / 764.8 | 675–722 / 737–779 |
| O4 | 239.3 / 254.4 | 315.7 / 344.1 | 31.2 / 32.2 | −12.2 / −14.4 | 401.4 / 411.0 | 419.2 / 435.4 | 381–438 / 397–454 |
| *Reserved* | | | | | | | |
| C3 | 418.1 / 468.8 | 481.0 / – | 31.3 / – | −54.5 / – | 551.1 / 604.7 | 564.0 / 616.1 | 531–580 / 584–632 |
| O3 | 468.6 / 486.1 | – | – | – | 608.5 / 619.6 | 604.5 / 625.6 | 575–621 / 596–641 |
| *Angular controls ($G_0$)* | | | | | | | |
| C1m | 461.5 | 490.2 | 60.0 | −79.6 | 548.1 | – | – |
| C3m | 250.9 | 303.1 | 47.8 | −52.2 | 383.7 | – | – |
| Om | 415.2 / 536.9 | – | – | – | 514.7 / 647.2 | – | – |

## S6. Numerical uncertainty

For the tested configurations, donor-induced differences are substantially less sensitive to numerical convergence settings than absolute zero-field-splitting parameters, because the vacancy and the complex share almost every numerical error. Raising the cutoff to 700 eV, sampling a 2 × 2 × 1 mesh and stretching the cell to 35 Å shift $D$ of $V_B^-$ by −8.37, +23.57 and +10.83 MHz but $\Delta D$ of C2 by only −0.17, −0.13 and +0.002 MHz. The combined uncertainty $u$ is the root-sum-square of the largest shift from each tested choice in $V_B^-$ and C2, transferred to the other placements: a rough estimate, neither a covariance nor a confidence interval. Absolute $D$ carries $u$ = 27.3 MHz, whereas $u(\Delta D)$ and $u(\Delta A_{zz})$ are at most 0.28 and 0.024 MHz.

C2 is the only tested system with nonzero $|E|$, so we assume shifts proportional to $|E|$. Then $u(|E|)$ is about 0.17% of $|E|$, 0.40–2.40 MHz over the 12 calibration tensors, and the estimate $u(\theta_E) = u(|E|)/(2|E|)$ is 0.047–0.048°, about 14 times the largest measured rotation. This uncertainty covers numerical choices within PBE only; cell size is a separate systematic (Sec. S5).

## S7. Relaxation, strain and hyperfine response

The lattice distortion drives most of the relaxed $\Delta D$ and little of $|E|$. Geometry-only calculations, with the bare charged vacancy in the relaxed C2 or O2 geometry, split the path of any observable $O$ from vacancy to relaxed complex $X$ into

$$\Delta O_{\mathrm{fixed}} = O_X(G_0) - O_{V_B}(G_0), \quad \Delta O_{\mathrm{geo}} = O_{V_B}(G_X) - O_{V_B}(G_0), \tag{S3}$$

with total $\Delta O_{\mathrm{tot}} = O_X(G_X) - O_{V_B}(G_0)$ and interaction term $\delta O = \Delta O_{\mathrm{tot}} - \Delta O_{\mathrm{fixed}} - \Delta O_{\mathrm{geo}}$. For C2 the frozen-lattice, geometry-only and total terms are $\Delta D$ = −54.1, −98.6 and −118.7 MHz with $|E|$ = 759.9, 93.4 and 735.7 MHz; for O2, −15.9, −28.6 and −32.6 MHz with $|E|$ = 581.0, 66.3 and 637.6 MHz (Fig. S3(c)). The distortion gives 83% and 88% of $\Delta D$ but only 13% and 10% of $|E|$, for two placements and one path; with the donor placed first, the lattice gets 54% and 51% of $\Delta D$. The interaction term $\delta D$, +34.0 and +11.9 MHz, is below five times the cell-size change, so it is indicated, not established. No term of this decomposition is purely electrostatic or chemical. Relaxation changes $D$ by up to −81.9 MHz (O1) and reduces $|\Delta D|$ at C1, C3 and C1m.

Strain. Six clamped-ion calculations apply ±0.1% biaxial, deviatoric and shear strain to the relaxed $V_B^-$ (Fig. S3(d)). Biaxial strain gives $b = \partial D/\partial \varepsilon_b$ = −12.06 GHz per unit strain and leaves $E$ unchanged to 0.001 MHz. Deviatoric and shear strain give 19.21 and 19.20 GHz per unit strain in $E$, the one coupling $s$ that trigonal symmetry requires. At the $G_X$ strain of the nitrogen triangle, $b$ and $s$ reproduce the geometry-only $\Delta D$ of C2 (−94.1 against −98.6 MHz) but not of O2 (−9.9 against −28.6 MHz) and overestimate both geometry-only $|E|$ (201.5 and 151.6 MHz).

Hyperfine response. The $V_B^-$ hyperfine tensors have been measured [53]. Core polarization [40] lowers their out-of-plane $A_{zz}$ in PBE from 48.3 to 35.0 MHz, away from the measured 47 MHz [1]. In the frozen lattice C1m lowers $A_{zz}$ on all three nitrogens, whereas every calibration donor raises it on its nearest nitrogen N30, by up to +4.31 MHz (Fig. S3(a)). N30 loses mostly in-plane $p$ spin, whose dipolar and core-polarization contributions are negative, so $A_{zz}$ rises; without core polarization the rise is at most +0.85 MHz (Fig. S3(b)). Over the 18 calibration values, $\Delta A_{zz}$ tracks each nitrogen's change of spin moment ($r$ = −0.990).

Hyperfine harmonic. In the frozen lattice $E_+ \approx 222 \sum_j \Delta A_{zz,j} \exp(2i\beta_j)$, with the coefficient from all eleven tensors ($R^2$ = 0.947). The relation holds over the 0–30° sector ($R^2$ = 0.984) but not at the angular controls: at C1m the harmonic of the shifts is 0.40 MHz, giving 90 MHz against 461.5 MHz, and at C3m and Om it misses the axis by 12°. After relaxation it does not carry over: a separate fit gives a slope of magnitude 137 ($R^2$ = 0.819). Each $V_B^-$ site has quadrupole coupling $C_q$ = 1.781 MHz and asymmetry $\eta$ = 0.208.

(a) $\Delta A_{zz}$ (MHz), with core polarization

| | $R(G_0)$ (Å) | ○ $G_0$ near | ○ $G_0$ middle | ○ $G_0$ far | ● $G_X$ near | ● $G_X$ middle | ● $G_X$ far |
|---|---|---|---|---|---|---|---|
| ● C1† | 4.35 | 3.01 | −1.06 | −1.06 | 4.52 | −3.27 | −3.27 |
| ▲ C1m† | 4.35 | −0.97 | −0.97 | −1.37 | 3.08 | 3.08 | −5.49 |
| ● C2 | 5.02 | 3.00 | −0.56 | −0.92 | 0.07 | −0.11 | −4.01 |
| ● C3* | 6.65 | 1.95 | −0.27 | −0.22 | 2.02 | −0.86 | −1.38 |
| ▲ C3m | 6.65 | 0.69 | −0.15 | −0.32 | 1.01 | 0.42 | −2.44 |
| ● C4 | 7.54 | 1.51 | −0.11 | 0.05 | 0.44 | 0.19 | −1.53 |
| ■ O1 | 3.85 | 4.31 | −0.98 | −1.66 | 0.26 | −3.66 | −5.84 |
| ▲ Om | 5.24 | 0.88 | −0.60 | −0.75 | 2.36 | 0.91 | −3.14 |
| ■ O2† | 5.81 | 2.48 | −0.40 | −0.40 | 2.27 | −1.57 | −1.57 |
| ■ O3* | 6.33 | 2.29 | −0.29 | −0.28 | 1.74 | −0.42 | −1.83 |
| ■ O4 | 8.08 | 1.45 | −0.01 | 0.15 | 1.26 | −0.33 | −0.48 |

(b) $\Delta A_{zz}$ (MHz), without core polarization

| | ○ $G_0$ near | ○ $G_0$ middle | ○ $G_0$ far | ● $G_X$ near | ● $G_X$ middle | ● $G_X$ far |
|---|---|---|---|---|---|---|
| ● C1† | 0.12 | 0.29 | 0.29 | 0.98 | −1.54 | −1.54 |
| ▲ C1m† | −1.55 | −1.55 | 0.02 | 2.00 | 2.00 | −3.24 |
| ● C2 | 0.76 | 0.16 | 0.34 | −1.99 | 0.26 | −2.40 |
| ● C3* | 0.76 | 0.29 | 0.44 | 0.63 | −0.26 | −0.57 |
| ▲ C3m | 0.10 | −0.17 | 0.37 | 0.25 | 0.22 | −1.40 |
| ● C4 | 0.77 | 0.15 | 0.51 | −0.29 | 0.27 | −0.89 |
| ■ O1 | 0.18 | 0.54 | 0.20 | −4.21 | −1.88 | −3.46 |
| ▲ Om | −0.12 | −0.59 | 0.27 | 1.01 | 0.73 | −1.62 |
| ■ O2† | 0.85 | 0.39 | 0.39 | 0.36 | −0.63 | −0.63 |
| ■ O3* | 0.96 | 0.20 | 0.47 | 0.31 | 0.01 | −0.86 |
| ■ O4 | 0.78 | 0.33 | 0.48 | 0.51 | 0.03 | −0.08 |

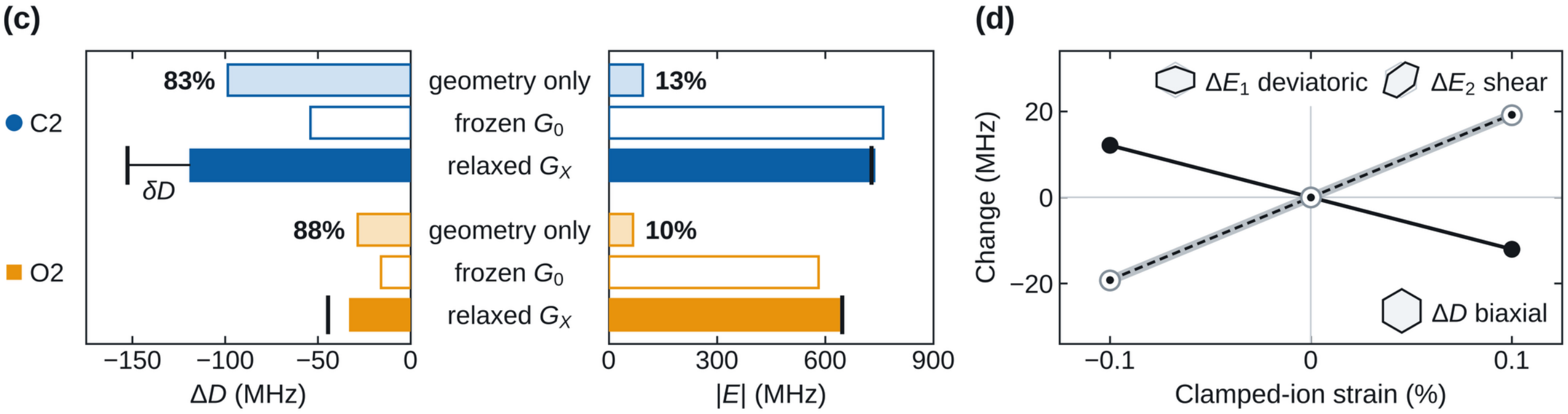

**Figure S3.** Nitrogen hyperfine changes, geometry-only decomposition and strain response. Uncorrected PBE, $\mathbf{8}\times\mathbf{8}$ cell. (a) Change of $A_{zz}$ on each vacancy nitrogen relative to $V_B^-$ (34.99 MHz), with core polarization; (b) without it (48.34 MHz). Rows follow chemistry, then distance; symbols as in Fig. 3; † marks two equidistant nitrogens. Red rise, blue fall; uncolored cells lie within the largest $\mathbf{8}\times\mathbf{8}\rightarrow\mathbf{10}\times\mathbf{10}$ change, 0.42 MHz in (a), 0.26 MHz in (b). (c) Decomposition of Eq. (S3) for C2 and O2; percentages are path-dependent geometry-only shares of relaxed values, ticks the frozen plus geometry-only sum, and the bracket the interaction term $\delta D$ (+34.0 MHz, C2). (d) Bare-vacancy clamped-ion response (hexagons, exaggerated modes), GHz per unit strain: $b = -12.06$; $s = 19.2$.

## S8. Functional and spin-treatment tests

Both tests treat $V_B^-$, C2 and O2 at both geometries and, for the angular contrast, C1 and C1m. They move $|E|$ in opposite directions, neither bounds the exact result, and we do not combine them.

HSE06. The hybrid functional [26,27] (25% screened exact exchange, screening parameter 0.2 Å$^{-1}$) was run as statics at the PBE geometries, with the uncorrected spin–spin sums over orbital pairs (pair sums); in the relaxed cells HSE06 forces reach 0.10–0.44 eV/Å. Of three criteria fixed before any HSE06 output was read, two hold. The C2–O2 $\Delta D$ contrast keeps its sign, going from −38.24 to −18.22 MHz at $G_0$ and from −86.16 to −74.07 MHz at $G_X$. The C2 axis turns by −0.14° and −0.38°, within tolerance, and O2 stays at 30.00°. The third criterion, $|E|$ within 15%, fails: HSE06 raises each C2 and O2 $|E|$ by 53.6–55.7%, with the C2/O2 ordering kept, so $|E|$ is quoted as a PBE value.

Outside the criteria, HSE06 lowers $D$ by 228–248 MHz in every $V_B^-$, C2 and O2 tensor. The near-donor N30 hyperfine change shrinks by 1.2–1.7 MHz and reverses sign for C2 at $G_X$, so it is functional-sensitive. The defect gap of these five tensors widens from 1.40–1.65 eV to 2.80–3.31 eV, and all keep the same two-hole triplet; O2 at $G_X$ passes the state check only with an energy window widened post hoc.

Spin decontamination. The triplets are nearly spin-pure: $\langle S^2\rangle$ is 2.00814–2.00883 (ideal 2) for $V_B^-$ and the calibration placements, a diagnostic that cannot show the pair sum to be uncontaminated. The $S = 1$ correction of Biktagirov et al. [28] uses the pair sums of PBE broken-symmetry determinants,

$$\tilde{D} = \frac{R_T - \langle R_{BS}\rangle}{2}, \qquad \text{(S4)}$$

where $R_T$ is the raw pair sum of the triplet and $\langle R_{BS} \rangle$ its average over the distinct $S_z = 0$ localizations. The released code returns no pair sum at $S_z = 0$, so we re-evaluated these determinants with a modified build of the same code version, which reproduces each term of the $V_B^-$ triplet pair sum to $6 \times 10^{-5}$ MHz. Decontamination raises $D$ of $V_B^-$, C2 and O2 by 325–363 MHz, to 3123.8 MHz for $V_B^-$. Under the HSE06 criteria the C2–O2 contrast keeps its sign, −20.42 and −86.59 MHz at $G_0$ and $G_X$, but the C2 axis turns by −1.59° and −5.11°, beyond tolerance. Post hoc, $|E|$ falls by 37–53% for C2 and O2. That fall and the $G_0$ contrast are indications, not corrections: the C2 average rests on a single stable localization, and one unaveraged determinant can carry an $|E|$ artifact of up to about 127 MHz.

Angular pair, HSE06. Criteria were fixed before the calculations started at $G_0$ and after they had started but before either had converged at $G_X$: the ratio $|E|(\mathrm{C1m})/|E|(\mathrm{C1})$ within ±5% of PBE, the axis branches, and the C1m–C1 $\Delta D$ difference relative to the $|E|$ change. The ratio is 0.432 (−7.6%) at $G_0$ and 0.561 (−6.5%) at $G_X$, outside ±5% but below the qualitative thresholds 0.734 and 0.800. The axes stay at 30.00° and 60.00°, and the $\Delta D$ difference is −15.95 and −43.05 MHz (PBE +2.03 and −2.22 MHz). The angular reduction persists under HSE06, although its magnitude differs from PBE: the ratio is quoted as a PBE value with the HSE06 shift as a functional uncertainty, and the statement that $\Delta D$ barely changes with azimuth is made for PBE only. HSE06 raises $|E|$ by 58.4% (C1) and 46.3% (C1m) at $G_0$ and by 55.6% and 45.4% at $G_X$.

Angular pair, decontamination. Criteria were fixed before any re-evaluation had run. C1 has one distinct localization at $G_0$ and two at $G_X$, and C1m one at $G_0$. All criteria hold at $G_0$: the ratio is 0.346, below 1 (0.793 if each $|E|$ moves against the contrast by the single-localization uncertainty, 126.59 MHz), the axes stay at 30.00° and 60.00°, and the $\Delta D$ difference keeps its sign at +36.69 MHz. Post hoc, $|E|$ falls by 49% (C1) and 62% (C1m). A single-localization average is an indication, not a correction: the two C1 localizations at $G_X$ differ by 43.3 MHz in $D$.

## S9. Field-plus-gradient model and its tests

This section defines the model of Sec. 6 and reports how its recorded predictions fared. The field $\mathbf{F} = -\nabla\varphi$ of a charge $+e$ at the donor is evaluated at the vacancy center, the centroid of the three vacancy nitrogens. Because $E_+$ rotates as $\exp(2\mathrm{i}\theta)$ and $F_+$ as $\exp(\mathrm{i}\theta)$, the threefold axis and mirrors require $E_+ = \mathrm{i}\kappa\bar{F}_+$ with real $\kappa$ at linear order; in-plane rank-two sources couple with real coefficients. $F_+$ and $G_+$ include the charge's in-plane periodic images and a uniform background, for the Rytova–Keldysh potential of a polarizable sheet [32],

$$\varphi(r) = \frac{e}{4\pi\varepsilon_0}\frac{\pi}{2r_0}\left[H_0(r/r_0) - Y_0(r/r_0)\right], \tag{S5}$$

with $H_0$ and $Y_0$ the Struve and second-kind Bessel functions. A least-squares fit of Eq. (6) to the 12 calibration tensors, about six independent constraints, gives $\kappa = -2959.6 \pm 97.8$ MHz per V Å$^{-1}$, $g_G = 3812.4 \pm 322.8$ MHz per V Å$^{-2}$ and $D_0 = -30.9 \pm 6.9$ MHz (leave-one-out jackknife errors 165.8, 418.6 and 9.9). With O1 and C1 in the calibration set, $\kappa$ and $g_G$ are effective parameters. A constant $D_0$ cannot be a property of the complex: the oxygen $G_0$ $\Delta D$ contains a cell-level component that must vanish in the isolated limit (Sec. S5). The clamped-ion $r_0 = 7.08$ Å comes from isolated-donor cells. The nonlinear variant adds a quadratic field term and uses the relaxed-ion value, 10.09 Å, which otherwise serves only as a sensitivity. The empirical $s_{\mathrm{geo}} = -8.155$ GHz per unit strain, from the two geometry-only calculations, stands in for the inhomogeneous relaxation field. The form was assembled, not selected blind, from four candidates whose leave-one-out results were known.

Interference. At $G_0$,

$$|E_+|^2 = \kappa^2|F_+|^2 + g_G^2|G_+|^2 + 2\kappa g_G\,\mathrm{Im}(F_+G_+)\,, \tag{S6}$$

where $\mathrm{Im}(F_+G_+) \propto -\sin 3\psi$ for a point charge. On the two mirror lines the terms are collinear, so $|E|(\mathrm{C1m})/|E|(\mathrm{C1}) = (1-x)/(1+x)$ with $x = g_\mathrm{G}|G_+|/(|\kappa||F_+|)$ at C1, and the C1m axis stays on the field-term branch while $x < 1$. The quadratic field term of the nonlinear variant interferes in the same way, so C1m and C3m test a rank-two term interfering with the field term, not the gradient itself.

Three-site expression. One real constant times the bare Coulomb potential at the three nitrogens, $r_\mathrm{N}$ from their centroid, with the charge's in-plane periodic images, gives $E_+ = c_\mathrm{V} S_\mathrm{V}$ with

$$S_\mathrm{V} = \sum_j \varphi\left(r_j\right)\exp\left(2\mathrm{i}\beta_j\right) = -\frac{3\mathrm{i}}{2}\, r_\mathrm{N}\bar{F}_+ + \frac{3}{4}\, r_\mathrm{N}^2 G_+ + O\left(r_\mathrm{N}^3\right), \tag{S7}$$

so $S_\mathrm{V}$ contains the field and gradient terms with fixed weights, in the ratio $r_\mathrm{N}/2$ = 0.76 Å, against the fitted $g_\mathrm{G}/|\kappa|$ = 1.29 Å. With $c_\mathrm{V}$ = 487.9 MHz per V, its $G_0$ leave-one-out mean absolute errors are 90.8 MHz in $|E|$ and 2.46° in $\theta_E$ over the six calibration placements.

Bare vacancy. With the ions at their relaxed positions, a modified build of the same code version applied a periodic in-plane potential, at wave vector 0.361 $\text{Å}^{-1}$, with fields up to ±0.20 V $\text{Å}^{-1}$ and $E'$ gradients of ±0.049 V $\text{Å}^{-2}$; it is not the potential of real external charges. Per unit DFT-screened field at the vacancy, about 0.35 of the applied one, the linear coefficient is 1.04 times the fitted $\kappa$ and $g_\mathrm{G}$ = 3123 MHz per V $\text{Å}^{-2}$, 0.82 times the fitted value (1.34 and 1.09 with the fitted couplings re-evaluated at the $r_0$ = 4.65 Å of that screening). The quadratic field response, by contrast, is 0.22 of the rank-two term at C1, so the rank-two term is mainly the vacancy's own response to the field gradient; a share lies within one standard deviation of the "mainly" threshold.

With no donor-fitted parameter, these couplings predict the frozen-lattice $E$ vectors of the 11 tensors beyond 5 Å to 55.9 MHz rms, against 54.1 MHz for the fitted model; on 6 other larger-cell tensors the error is 122.9 MHz against 88.3 MHz. Criteria were fixed before these runs. One leading-order check on the symmetry-forbidden term in $D$ linear in the field failed, 0.287 MHz against a limit of 0.107 MHz, and excludes no run.

Archived predictions. The predictions and evaluation criteria were archived together before the corresponding tensors were computed. The C3 and O3 predictions were fixed on 17 October 2026 at 12:04 UTC, before the first calculation started at 12:05 UTC; their $G_X$ values are conditional on relaxed geometries visible while the models were built. The C1m and C3m predictions at $G_0$ were fixed on 18 October at 02:21 UTC, before their calculations, and the O4 12 × 12 prediction, without criteria, at 09:58 UTC. Om followed on 23 October at 01:44 UTC. Predictions for the relaxed controls, fixed after each relaxation had passed its automated checks and before its tensor was computed, are less blind than the $G_0$ test.

Each $E$ vector has a joint circular region built from the leave-one-out errors, with 97.5% radii of 281.3 and 222.0 MHz at $G_0$ and $G_X$. The regions are wide: the $G_0$ radius is 0.61 of the predicted $|E|$ at C3 and O3 and exceeds it at C3m, where inclusion is uninformative. Simulated coverage is 0.966–0.969 for the $E$ vector but only 0.87–0.92 for the 95% $\theta_E$ intervals, and eight placements cannot verify the residual model behind these numbers.

Outcomes (Tables S3 and S4, Fig. S4). The predicted $E$ vectors for the reserved C3 and O3 configurations fall within the estimated uncertainty regions, but the model does not consistently outperform the comparison models in these cases. For the angular controls C1m and C3m it reproduces the direction-dependent response more successfully, with $E$-vector errors of 36.4 and 17.1 MHz, although this agreement does not uniquely establish a field-gradient mechanism. DFT gives the predicted C1m branch, but so does the field-only model, which shows only that the field term dominates, and the nonlinear variant gives frozen-lattice $|E|$ ratios of 0.463 at C1m and 0.755 at C3m, against 0.467 and 0.600 in DFT. The relaxed controls miss by 26.9 and 28.8 MHz, and Om by 41.4 MHz at $G_0$ and 37.9 MHz relaxed. With two carbon placements and one model form, this description of $|E|$ and $\theta_E$ is phenomenological, and it does not extend to $\Delta D$.

**Table S3.** Errors of the model, Eq. (6), and the comparison models (PBE, $\mathbf{8} \times \mathbf{8}$; MHz, degrees; prediction minus DFT; paired entries $\boldsymbol{G_0}$ / $\boldsymbol{G_X}$, C3 / O3 or C1m / C3m / Om; C1m / C3m at $\boldsymbol{G_X}$). $\boldsymbol{N_p}$: fitted parameters. Leave-one-out mean absolute errors $\langle|\boldsymbol{\delta E}|\rangle$, $\langle|\boldsymbol{\delta\theta_E}|\rangle$ (four off-mirror placements) and $\langle|\boldsymbol{\delta\Delta D}|\rangle$, and root-mean-square $\boldsymbol{E}$-vector error $\boldsymbol{\delta E_{rms}}$. Reserved, angular: recorded $\boldsymbol{E}$-vector error ($\boldsymbol{\theta_E}$ error). Power law: $|\boldsymbol{E}| \propto \boldsymbol{R^n}$ at the mean $\boldsymbol{\theta_E}$ of the calibration placements and $\boldsymbol{\Delta D}$. Constant angle: model $|\boldsymbol{E}|$ at that angle (relaxed: angular controls only; C3/O3 errors outside the recorded predictions). Nonlinear: terms selected on all data or per fit. C1m allows only 60° or −30° (0.00°: predicted branch).

| Model | $N_p$ | $\langle\lvert\delta E\rvert\rangle$, $G_0$ / $G_X$ | $\delta E_{rms}$, $G_0$ / $G_X$ | $\langle\lvert\delta\theta_E\rvert\rangle$ (°), $G_0$ / $G_X$ | $\langle\lvert\delta\Delta D\rvert\rangle$, $G_0$ / $G_X$ | Reserved, C3 / O3 | Angular, C1m / C3m / Om; C1m / C3m at $G_X$ |
|---|---|---|---|---|---|---|---|
| Model | 3 | 94.1 / 76.7 | 120.4 / 95.0 | 1.06 / 1.12 | 25.5 / 19.4 | 41.3 (+0.34°) / 18.1 (+0.97°) | 36.4 (0.00°) / 17.1 (+0.81°) / 41.4 (−0.23°); 26.9 (0.00°) / 28.8 (+1.88°) |
| Field-only model | 2 | 35.6 / 69.6 | 223.0 / 183.5 | 9.15 / 6.18 | 38.3 / 40.9 | 74.7 (+4.51°) / 147.9 (+9.14°) | 595.0 (0.00°) / 152.3 (+7.92°) / 320.6 (+8.55°); 363.0 (0.00°) / 82.4 (+5.42°) |
| Nonlinear (all data) | 3 | 72.1 / 49.4 | 121.5 / 82.1 | 4.11 / 1.75 | 24.4 / 18.8 | 41.6 (+2.33°) / 74.2 (+4.55°) | 82.2 (0.00°) / 96.9 (+5.16°) / 105.9 (+2.30°); 152.0 (0.00°) / 30.9 (+2.66°) |
| Nonlinear (per fit) | 3 | 103.3 / 63.7 | 145.2 / 96.9 | 4.13 / 3.55 | 35.7 / 29.9 | – | – |
| Power law | 2 | 42.0 / 67.3 | 59.4 / 120.3 | 1.83 / 3.63 | 27.2 / 35.8 | 37.3 (+0.36°) / 45.4 (−1.39°) | 879.4 (−28.95°) / 213.0 (−15.82°) / 354.2 (−13.48°); 900.6 (−26.98°) / 216.7 (−16.35°) |
| Constant angle | 1 | 94.1 / – | 121.7 / – | 1.83 / – | – | 41.3 (+0.36°) / 24.2 (−1.39°) | 465.5 (−28.95°) / 141.8 (−15.82°) / 207.2 (−13.48°); 644.1 (−26.98°) / 191.5 (−16.35°) |

**Table S4.** Recorded predictions against the DFT tensors computed afterward (PBE, $\mathbf{8} \times \mathbf{8}$ cell; MHz unless noted). $\boldsymbol{E}$-vector error: distance between predicted and DFT $(\boldsymbol{E_1}, \boldsymbol{E_2})$, and whether it lies inside the joint 97.5% circle (Sec. S9; Om, one placement: 95% circles, radii 243.9 MHz at $\boldsymbol{G_0}$ and 192.5 MHz at $\boldsymbol{G_X}$). $\boldsymbol{G_X}$ predictions are conditional on the relaxed geometries. At $\boldsymbol{G_0}$ the predicted $\boldsymbol{\Delta D}$ is the constant $\boldsymbol{D_0}$, so its closeness at Om is no evidence that the model describes $\boldsymbol{\Delta D}$.

| Tensor | $\lvert E\rvert$ predicted | $\lvert E\rvert$ DFT | $\theta_E$ predicted (°) | $\theta_E$ DFT (°) | $\Delta D$ predicted | $\Delta D$ DFT | $E$-vector error (circle) |
|---|---|---|---|---|---|---|---|
| *Reserved placements* | | | | | | | |
| C3 at $G_0$ | 459.0 | 418.1 | 31.03 | 30.69 | −30.9 | −51.7 | 41.3 (inside) |
| O3 at $G_0$ | 459.6 | 468.6 | 33.41 | 32.44 | −30.9 | −16.7 | 18.1 (inside) |
| C3 at $G_X$ (conditional) | 504.5 | 468.8 | 31.14 | 32.76 | −37.3 | −41.5 | 45.1 (inside) |
| O3 at $G_X$ (conditional) | 504.5 | 486.1 | 34.38 | 35.20 | −47.1 | −36.5 | 23.2 (inside) |
| *Angular controls* | | | | | | | |
| C1m at $G_0$ | 497.9 | 461.5 | 60.00 | 60.00 | −30.9 | −78.4 | 36.4 (inside) |
| C3m at $G_0$ | 266.4 | 250.9 | 47.68 | 46.87 | −30.9 | −48.2 | 17.1 (inside) |

| Tensor | $\lvert E \rvert$ predicted | $\lvert E \rvert$ DFT | $\theta_E$ predicted (°) | $\theta_E$ DFT (°) | $\Delta D$ predicted | $\Delta D$ DFT | $E$-vector error (circle) |
|---|---|---|---|---|---|---|---|
| C1m at $G_X$ (conditional) | 722.9 | 696.0 | 60.00 | 60.00 | −16.6 | −52.5 | 26.9 (inside) |
| C3m at $G_X$ (conditional) | 348.0 | 329.6 | 51.25 | 49.37 | −60.5 | −55.4 | 28.8 (inside) |
| *Oxygen angular control* | | | | | | | |
| Om at $G_0$ | 456.4 | 415.2 | 44.31 | 44.54 | −30.9 | −27.1 | 41.4 (inside) |
| Om at $G_X$ (conditional) | 539.5 | 536.9 | 45.70 | 47.71 | −32.7 | −31.0 | 37.9 (inside) |

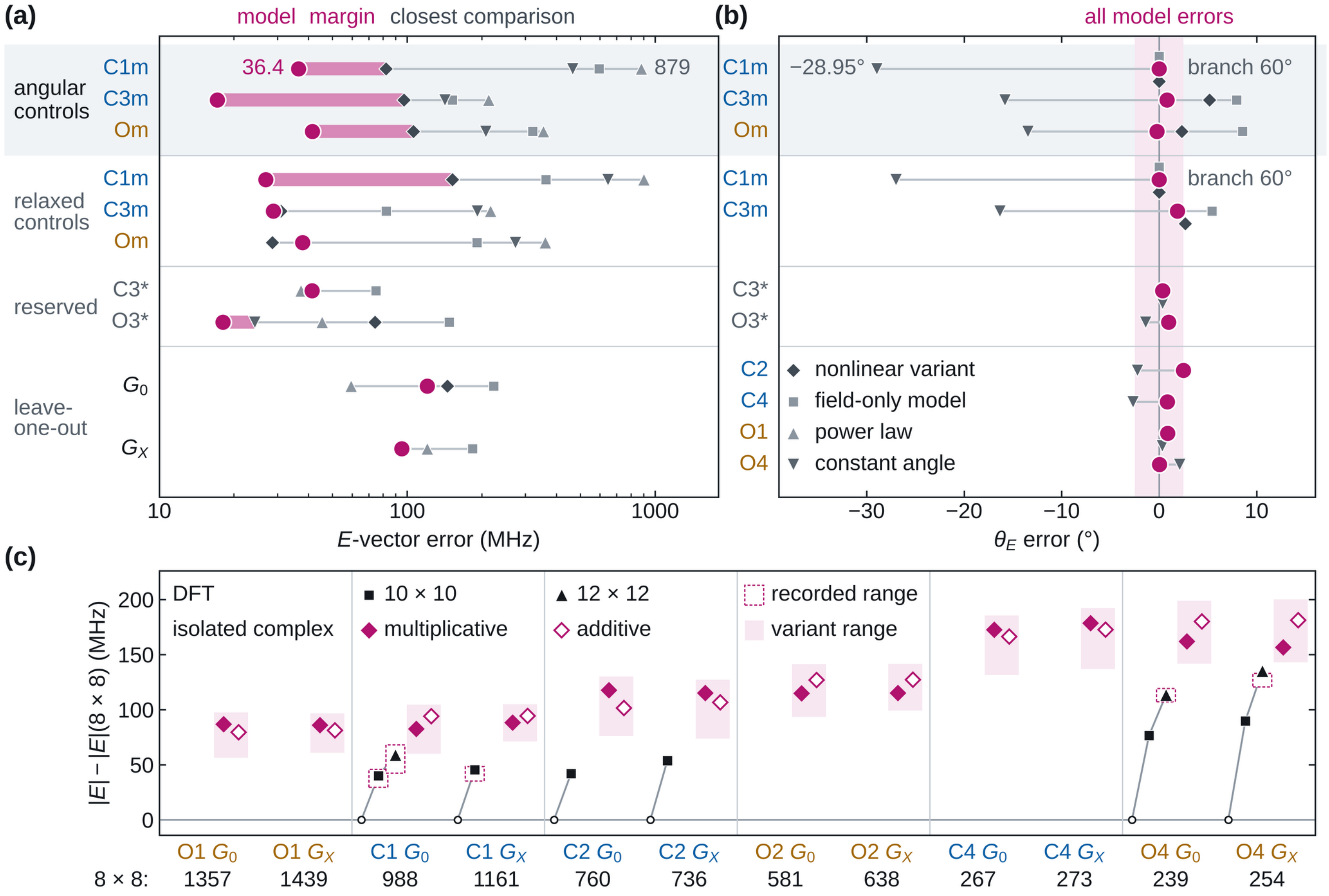

**Figure S4.** Prediction errors of the screened field-plus-gradient model against uncorrected PBE. (a) $\boldsymbol{E}$-vector errors of the model and four comparison models; the field-only model (unscreened) takes $\lvert \boldsymbol{E} \rvert$ from the field strength, the power law from distance. Light magenta bars: the model's margin over the closest comparison model. Predictions for angular controls, relaxed controls and reserved placements were recorded before the calculation, at $\boldsymbol{G_0}$ or given the relaxed structure, except the constant-angle errors at C3* and O3*. Leave-one-out: rms errors over six calibration placements. (b) $\boldsymbol{\theta_E}$ errors, prediction minus DFT; leave-one-out off-mirror at $\boldsymbol{G_0}$; power law: constant angle. Branch 60°: C1m axis across the mirror for model, field-only model, DFT. (c) $\lvert \boldsymbol{E} \rvert$ minus the periodic $\mathbf{8 \times 8}$ value: DFT $\mathbf{10 \times 10}$ and $\mathbf{12 \times 12}$, isolated-complex model estimates with variant ranges; dashed boxes, ranges recorded beforehand (relaxed C1: five variants).

## S10. Simulated spectra

Under the simulated conditions, no configuration pair is distinguishable when the transverse offset is unrestricted. Spectra (Fig. S5(a)) use the 81-state Hamiltonian, Eq. (S1), with $g = 2.0023$, at zero field and 5, 20 or 50 mT along the normal or tilted by 30°. Gaussian hyperfine lines have FWHM 95 MHz, reported for natural-abundance hBN [33], or 44 MHz, reported for $h^{10}B^{15}N$ [54], hypothetical for $^{14}N$; tilted fields average three orientations. Model spectra use isolated-complex estimates (Sec. S5), and no optical-readout contrast is modeled.

For data $X$ and model $Y$ the penalized spectral distance is

$$S_{\mathrm{pen}}^2 = \min\left[\chi_{\mathrm{data}}^2 + \frac{|\delta E_+|^2}{\sigma_{\mathrm{strain}}^2 + \sigma_{\mathrm{th},X}^2 + \sigma_{\mathrm{th},Y}^2}\right], \tag{S8}$$

with $\chi_{\mathrm{data}}^2$ the noise-weighted misfit minimized over $\delta E_+$, field and hyperfine scales, the linewidth and a common $D$ offset, so $\Delta D$ adds nothing. The prior term uses $\sigma_{\mathrm{strain}}$ = 50 MHz, the ensemble $|E|$ [1], an assumed donor-independent distribution that also contains charge contributions; the theory width $\sigma_{\mathrm{th}}$ (12.0–44.5 MHz) covers the 10 × 10 image residual, not the spread of the isolated-complex extrapolation. $S_{\mathrm{pen}}$ is a distance between noiseless spectra [55], not a $p$ value or an assignment probability. A pair is separated when $S_{\mathrm{pen}} \geq 5$ and five times the finite-size floor, the largest $S_{\mathrm{pen}}$ between cell sizes, both declared exploratory conventions.

Monte Carlo. Each realization draws an offset from the prior and white noise (SNR, the zero-field peak-to-noise ratio per MHz), refits both model spectra from data-derived starts and decides for the smaller penalized $\chi^2$. We ran 40 800 realizations in 102 sets of conditions for five chosen pairs (Fig. S5(c)). Mean misassignment rates of carbon–oxygen pairs follow $S_{\mathrm{pen}}$ (C1/O1 at 95 MHz and SNR 300: 5.12 gives 0.6% [0.2, 1.5]).

What separates pairs. The prior does most of the work: at SNR 100 it supplies a median 99.7% of $S_{\mathrm{pen}}^2$ at the best setting (95 MHz), and with $E$ unconstrained no pair reaches 5 (largest 2.02, at 44 MHz). Under the ensemble prior 24 of 36 pairs (nine configurations) are separated at both linewidths; periodic 8 × 8 tensors with their finite-size width separate 10 at 95 MHz. At SNR 100 and the PBE $|E|$ scale no matched carbon–oxygen pair reaches 5 in any tensor or prior variant (largest 4.98). With three $^{15}N$ at 44 MHz, C1/O1 reaches 6.04, the only carbon–oxygen case separated by the spectrum rather than the prior, and only within the 8 × 8 model. C1 and O1 also differ in distance and azimuth, so this is not a chemical contrast.

Twelve-configuration test. Spectra of all twelve relaxed configurations were assigned at one set of conditions fixed in advance: 50 mT tilted 30°, SNR 100, polarized populations, $^{14}N$ and 400 realizations per configuration (Fig. S5(b)). The angular-control model spectra were extrapolated after their tensors were known. The mean correct-assignment rate (normal-approximation 95% interval) is 0.851 [0.842, 0.861] at 95 MHz and 0.818 [0.808, 0.828] at 44 MHz. Of donor spectra at 95 MHz, 16.2% [15.1, 17.3] are assigned to another distance; the rate is 1.8% [0.7, 3.6] for C1m and 100.0% [99.1, 100.0] for C1m and C3m without the angular-control model spectra (95% Clopper–Pearson intervals [56]).

Limits. These rates are not experimental accuracies: they hold within this set of model spectra and assume the generating offset distribution and one known factor between computed and measured $|E|$. On the first 200 realizations per configuration at 95 MHz, offsets of 100 MHz width change the mean correct-assignment rate by −0.212. Scaling every donor $|E|$ by 1.55 or 0.5 gives 0.938 or 0.575, against 0.857 unscaled on the same draws, and spectra at HSE06/PBE $|E|$ ratios scored with PBE model spectra fall to 0.217. Transferring the monolayer $|E|$ to shallow vacancies in thicker flakes [34] or on substrates needs a model of nonlocal screening [32], and optical readout may change the computed charge configuration (Secs. S3 and S8).

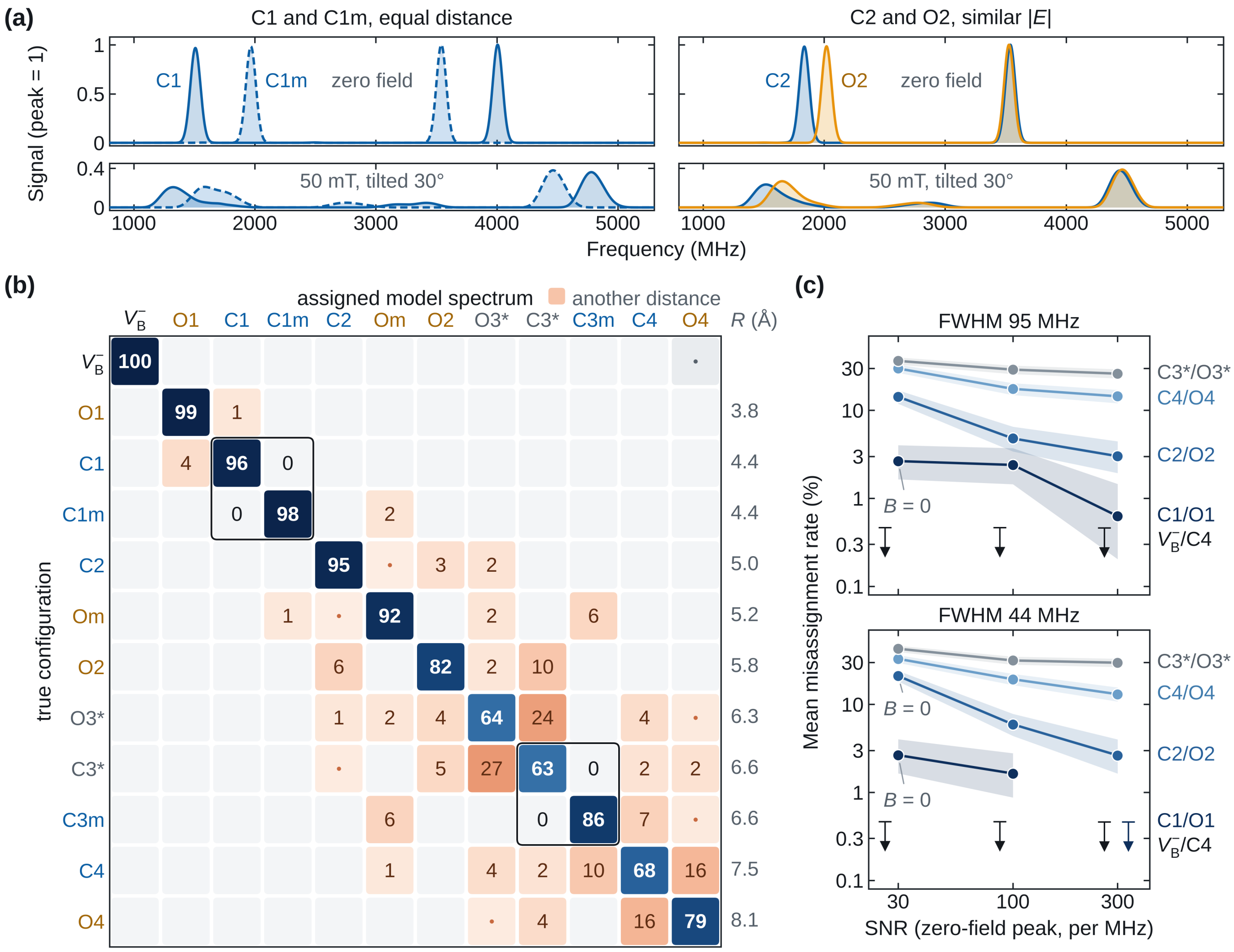

**Figure S5.** Constraints from simulated spectra. Relaxed isolated-complex tensors (model estimates), uncorrected PBE, three $^{14}\mathbf{N}$, $\boldsymbol{\sigma}_{\mathbf{strain}}$ = 50 MHz, 95 MHz lines unless marked. (a) Noiseless spectra of C1 and C1m (dashed), at equal distance, and of C2 and O2, whose $|\boldsymbol{E}|$ bracket that of C1m, at zero field and in 50 mT tilted 30°. (b) Percentage of spectra of each configuration (rows) assigned to each model spectrum (columns) at SNR 100 and tilted 50 mT, ordered by ideal-lattice $\boldsymbol{R}$. Blue, correct assignments; warm tiles, assignments to a donor at another distance; dots, below 1%; outlined, equal-distance pairs. (c) Mean misassignment rate (worse population model) against SNR with 95% Clopper–Pearson intervals, about 800 realizations per pair, tilted 50 mT, or $\boldsymbol{B} = \mathbf{0}$ where marked and for every $V_B^-$/C4 bound; arrows, upper bounds.

## S11. Electronic structure, spin-imbalance relation and map

This section gives the details behind Secs. 3 and 5 and the map and ensemble of Sec. 6, all exploratory. Densities of states and unfolded bands come from PBE calculations on the production densities of the $G_0$ tensors, on a $3 \times 3 \times 1$ mesh; for $V_B^-$ five further points of zero weight complete the primitive path Γ–M–K–Γ. Band unfolding follows Refs. [30] and [57], with the pseudo-wavefunction weight. States outside the sheet are omitted, because the net charge of the bare-vacancy cell artificially pulls vacuum states down by about 1.2 eV. At K the valence-band edge resonates with the filled $p_z$ levels, so we place it at the centroid of its spectral weight. HSE06 statics hold no K point, so their edges, which set the 5.97 eV of Sec. 3 and the HSE06 column of Fig. 2(c), come from the host states folded to Γ from the points nearest K, shifted to K by the offsets of the same states in the PBE $V_B^-$ run.

Levels. At each Γ-point tensor the hole levels and the filled $a_1'$ partner are the two empty and the one occupied minority levels with the largest in-plane $p$ weight on the three vacancy nitrogens. For a far carbon donor an empty carbon level can lie between or below the hole levels, so selection by energy alone would fail (Table S6, Fig. S6). Each hole orbital of the bare vacancy keeps 54% of its norm as PAW-projected weight on the three nitrogens. The DFT $n_j$ is the two hole orbitals' summed PAW-projected weight on nitrogen $j$, all orbital channels, divided by that one bare-vacancy share, 0.5385, and each tensor's own pair $D$ serves as $D_{\mathrm{h}}$ in the test of Eq. (2). $c_n$ is the modulus of the least-squares complex slope of $E_+$ on $\sum_j n_j \exp(2\mathrm{i}\beta_j)$ over the 22 Γ-point tensors.

Hole-pair tensor. For two orthonormal orbitals $\psi_1$, $\psi_2$ the dipolar tensor is

$$D_{\alpha\beta} = \frac{1}{2}\frac{\mu_0}{4\pi}\frac{g^2\mu_{\mathrm{B}}^2}{h}\frac{4\pi}{\Omega}\sum_{\mathbf{G}\neq 0}\left(\frac{G_\alpha G_\beta}{G^2} - \frac{\delta_{\alpha\beta}}{3}\right)\left[\rho_{11}(\mathbf{G})\rho_{22}^*(\mathbf{G}) - |\rho_{12}(\mathbf{G})|^2\right], \quad \text{(S9)}$$

with $\rho_{ab}(\mathbf{G})$ the Fourier transform of $\psi_a^*\psi_b$ and $\Omega$ the cell volume, the two-orbital term of the pair sum of Ref. [23], evaluated from pseudo-wavefunctions rescaled to unit norm, without one-center terms. Closed shells cancel in the full pair sum when majority and minority orbitals coincide [58], which is why two orbitals can carry $D$. They do not coincide in the unrestricted determinant: the pair gives $D$ within 3.5% of the full value and $|E|$ at 0.45–0.56 of it (Table S6).

Derivation of Eq. (2). Let the three orthonormal site orbitals be $|j\rangle$. Write the filled minority level as $|a\rangle$ and the two hole orbitals as $|x\rangle$, $|y\rangle$, so that the $3 \times 3$ matrix with rows $(x_j, y_j, a_j)$ is orthogonal. The antisymmetrized hole pair has amplitude $x_k y_l - y_k x_l$ on sites $k$ and $l$. This is a cofactor of the matrix, and in an orthogonal matrix a cofactor equals, up to sign, the element in the remaining position, so the pair weight on $(k, l)$ is $a_j^2$, with $j$ the third site. Neglecting overlap densities and assuming the point-dipole form, each pair contributes $\tau(1 - 3\hat{\mathbf{u}}\hat{\mathbf{u}})$ with $\hat{\mathbf{u}}$ along the edge, and the three edges point along $\beta_j + 90°$. Summing gives $D_{zz} = \tau\sum_j a_j^2$ and $E_+ = \frac{3}{2}\tau\sum_j a_j^2 \exp(2\mathrm{i}\beta_j)$. Since $\sum_j a_j^2 = 1$, $D = \frac{3}{2}\tau$ is independent of the distribution, and with $\delta(a_j^2) = -\delta n_j$ Eq. (2) follows. Edge lengths enter only through $\tau$, so this holds at a fixed triangle; a distortion changes $D$ directly.

The relation. Mirror symmetry requires $c$ to be real, and Eq. (2) gives its sign. Table S5 gives each tensor set; the strained vacancy has moments near the printed resolution, 0.001 $\mu_{\mathrm{B}}$ or about 4 MHz rms in $E_+$, and in the geometry-only tensors the distortion acts on the triangle directly. The constant belongs to the PAW-projected moment, so the dimensionless $c\,m_0/D$ = 1.26 in PBE (Table II) transfers more safely, though it is 1.67 in HSE06 statics. For frozen-lattice C1 the 58.4% rise of $|E|$ in HSE06 statics comes mainly from the imbalance: $|S_m|$ grows by a factor of 1.42 and $|E|/|S_m|$ by 1.11. In the first-order model the wider gap alone would lower the imbalance by a factor of 0.70, so exact exchange enhances the response beyond a simple inverse gap. Post hoc, spin decontamination removes a share of $|E|$ comparable to the remainder beyond the hole pair (Sec. S8), which suggests that much of this remainder is a spin-polarization property of the unrestricted determinant; the two are not the same quantity.

Three-site model. The lowest eigenvector of Eq. (4) gives $a_j$, with $t$ = 0.772 eV, $r_{\mathrm{N}}$ = 1.5292 Å and $V_j = -14.3996\ \mathrm{eV\,Å}/(\varepsilon r_j)$ for a charge $+e$, without images. With $c_n$ taken from the six frozen-lattice calibration tensors (Eq. (5)), 5.41 GHz per hole, $\varepsilon$ is scanned for the least rms $E$-vector error, 101 MHz. Over all eleven, the 5.27 GHz per hole of Eq. (3) gives 83 MHz rms; a free $c_n$ gives 67 MHz but an amplitude inconsistent with Eq. (3), so we keep it fixed. The first-order coefficient is $2c_n \times 14.3996\ \mathrm{eV\,Å}/(3\Delta\varepsilon)$ = 6.1 GHz Å, close to the $c_{\mathrm{V}}$ of Sec. S9, 7.0 GHz Å. Equation (5) gives the field coupling in closed form, $|\kappa| = c_n r_{\mathrm{N}}/\Delta$ = 3.48 GHz per V Å$^{-1}$ of local field, comparable to the fitted 2.96 and the bare 3.08. Divided by the response per unit applied field, 1.09 GHz per V Å$^{-1}$, 3.48 gives the $\varepsilon \approx$ 3.2 of Sec. 5.

The total self-consistent potential at the nitrogen cores, unlike the external one, does not give $E_+$ as a second harmonic ($R^2$ = 0.70 at $G_0$). The ratio of $|E|$ to the $e'$ splitting is 2.07–2.58 GHz per eV in the 0–30° sector and 1.56–1.89 at the angular controls, a separation the exact solution shares. The hole pair alone gives C1m/C1 = 0.523, against 0.467 for the full tensor.

Map and ensemble. The map evaluates Eq. (4) for a charge $+e$ on every lattice site beyond 3 Å from the vacancy center, at ideal positions. Inverted through the map, the nearest model vector returns the computed site for 7 of the 11 placements, all six within 6 Å and one of the five beyond. Far from the vacancy (Sec. 6), $|E| \approx (\mathcal{A}/R^2)[1 + (3r_{\rm N}/4R)\ \sin\ 3\psi]$, so the contrast between directions fades slowly.

The ensemble draws charges of both signs in equal numbers on distinct lattice sites beyond 3 Å, adds their potentials at the three nitrogens and solves Eq. (4) once per draw. At low density $E_+$ becomes a two-dimensional Cauchy variable of scale $\pi n \mathcal{A}$, whose median $|E|$ is 7.6 MHz per $10^{12}$ cm$^{-2}$, against 7.5 in the simulation. For charge noise one charge at distance $R$ switches off or hops to one of its six same-sublattice neighbors, 2.513 Å away. With $\varepsilon$ between 2.5 and 7, or a Rytova–Keldysh potential with $r_0$ = 11.5 Å (fitted at fixed $c_n$, 137 MHz rms), the density of the 50 MHz median runs from 2.3 × $10^{12}$ to 1.4 × $10^{13}$ cm$^{-2}$. The bias of these estimates shows in the $|E|$ of O4: 239 MHz at 8 × 8, 353 at 12 × 12 and 253 in the model.

**Table S5.** The relation $\boldsymbol{E_+ = -c\,S_m}$ by tensor set, with $\boldsymbol{c}$ the least-squares complex slope. Interval: 95% bootstrap over tensors. Phase: argument of the slope, 180° for a real negative constant. The 36 PBE tensors of the first four rows together give $\boldsymbol{c}$ = 7.95 (7.85–8.05) GHz per $\boldsymbol{\mu_B}$, $\boldsymbol{R^2}$ = 0.9987 and 25 MHz rms. Every $\boldsymbol{R^2}$ in this paper is uncentered, since its fit has no intercept; at the same slope the centered value for the 36 PBE tensors is 0.9959.

| Tensor set | $n$ | $c$ (GHz $\mu_{\rm B}^{-1}$) | Interval | Phase (°) | $R^2$ | rms (MHz) |
|---|---|---|---|---|---|---|
| PBE, 8 × 8, $G_0$ | 11 | 8.14 | 8.00–8.27 | 178.8 | 0.9989 | 21.6 |
| PBE, 8 × 8, $G_X$ | 11 | 7.80 | 7.69–7.95 | 180.0 | 0.9990 | 22.7 |
| PBE, 10 × 10 | 10 | 7.96 | 7.87–8.06 | 179.3 | 0.9992 | 20.4 |
| PBE, 12 × 12 | 4 | 7.96 | 7.73–8.33 | 180.0 | 0.9994 | 15.5 |
| HSE06 statics, 8 × 8 | 8 | 8.92 | 8.82–9.05 | 179.6 | 0.9995 | 27.1 |
| Bare vacancy, applied fields and gradients | 8 | 8.31 | 8.24–8.53 | 180.0 | 0.9991 | 4.1 |
| Bare vacancy, clamped-ion strain | 6 | 10.24 | 9.59–11.09 | 180.0 | 0.9949 | 1.1 |
| Bare vacancy, geometry only | 2 | 6.29 | 5.53–7.06 | 191.2 | 0.9618 | 15.8 |

**Table S6.** Hole levels and hole-pair tensors at $\boldsymbol{\Gamma}$, uncorrected PBE, $\boldsymbol{8 \times 8}$ cell. $\boldsymbol{\delta_{e'}}$: splitting of the $\boldsymbol{e'}$ pair; $\boldsymbol{\Delta}$: mean $\boldsymbol{e'}$ level minus $\boldsymbol{a_1'}$ level, minority spin; $|\boldsymbol{S_m}|$: spin imbalance; pair columns: Eq. (S9) for the two hole orbitals. The bare vacancy has $\boldsymbol{\Delta}$ = 2.315 eV and pair $\boldsymbol{D}$ = 2817.0 MHz, against a full $\boldsymbol{D}$ of 2798.8 MHz.

| Tensor | $\lvert E\rvert$ (MHz) | Pair $\lvert E\rvert$ | Pair $D$ | $\delta_{e'}$ (eV) | Δ (eV) | $\lvert S_m\rvert$ ($\mu_{\rm B}$) |
|---|---|---|---|---|---|---|
| | | | $G_0$ | | | |
| C1 | 987.6 | 494.6 | 2811.9 | 0.412 | 2.372 | 0.123 |
| C2 | 759.9 | 364.8 | 2800.2 | 0.328 | 2.341 | 0.094 |
| C4 | 267.4 | 125.9 | 2819.7 | 0.125 | 2.318 | 0.031 |
| O1 | 1357.4 | 630.0 | 2784.2 | 0.525 | 2.381 | 0.166 |
| O2 | 581.0 | 276.9 | 2821.6 | 0.256 | 2.335 | 0.068 |
| O4 | 239.3 | 114.1 | 2824.0 | 0.114 | 2.319 | 0.028 |
| C3 | 418.1 | 213.4 | 2822.0 | 0.202 | 2.326 | 0.052 |
| O3 | 468.6 | 220.6 | 2811.6 | 0.211 | 2.325 | 0.055 |
| C1m | 461.5 | 258.4 | 2816.3 | 0.295 | 2.327 | 0.061 |
| C3m | 250.9 | 139.4 | 2819.8 | 0.145 | 2.317 | 0.033 |
| Om | 415.2 | 209.6 | 2816.6 | 0.220 | 2.322 | 0.051 |
| | | | $G_X$ | | | |
| C1 | 1160.5 | 525.3 | 2787.4 | 0.329 | 2.407 | 0.147 |
| C2 | 735.7 | 346.4 | 2724.2 | 0.297 | 2.274 | 0.092 |
| C4 | 273.2 | 127.2 | 2776.6 | 0.124 | 2.283 | 0.033 |
| O1 | 1439.4 | 642.4 | 2669.3 | 0.433 | 2.351 | 0.189 |
| O2 | 637.6 | 288.6 | 2793.7 | 0.198 | 2.330 | 0.080 |

| Tensor | $\lvert E \rvert$ (MHz) | Pair $\lvert E \rvert$ | Pair $D$ | $\delta_{e'}$ (eV) | $\Delta$ (eV) | $\lvert S_m \rvert$ ($\mu_B$) |
|---|---|---|---|---|---|---|
| O4 | 254.4 | 116.7 | 2816.8 | 0.085 | 2.318 | 0.032 |
| C3 | 468.8 | 215.8 | 2805.7 | 0.156 | 2.322 | 0.059 |
| O3 | 486.1 | 222.2 | 2789.4 | 0.161 | 2.315 | 0.061 |
| C1m | 696.0 | 345.7 | 2797.6 | 0.111 | 2.384 | 0.089 |
| C3m | 329.6 | 160.9 | 2788.8 | 0.077 | 2.303 | 0.044 |
| Om | 536.9 | 252.8 | 2805.6 | 0.162 | 2.336 | 0.068 |

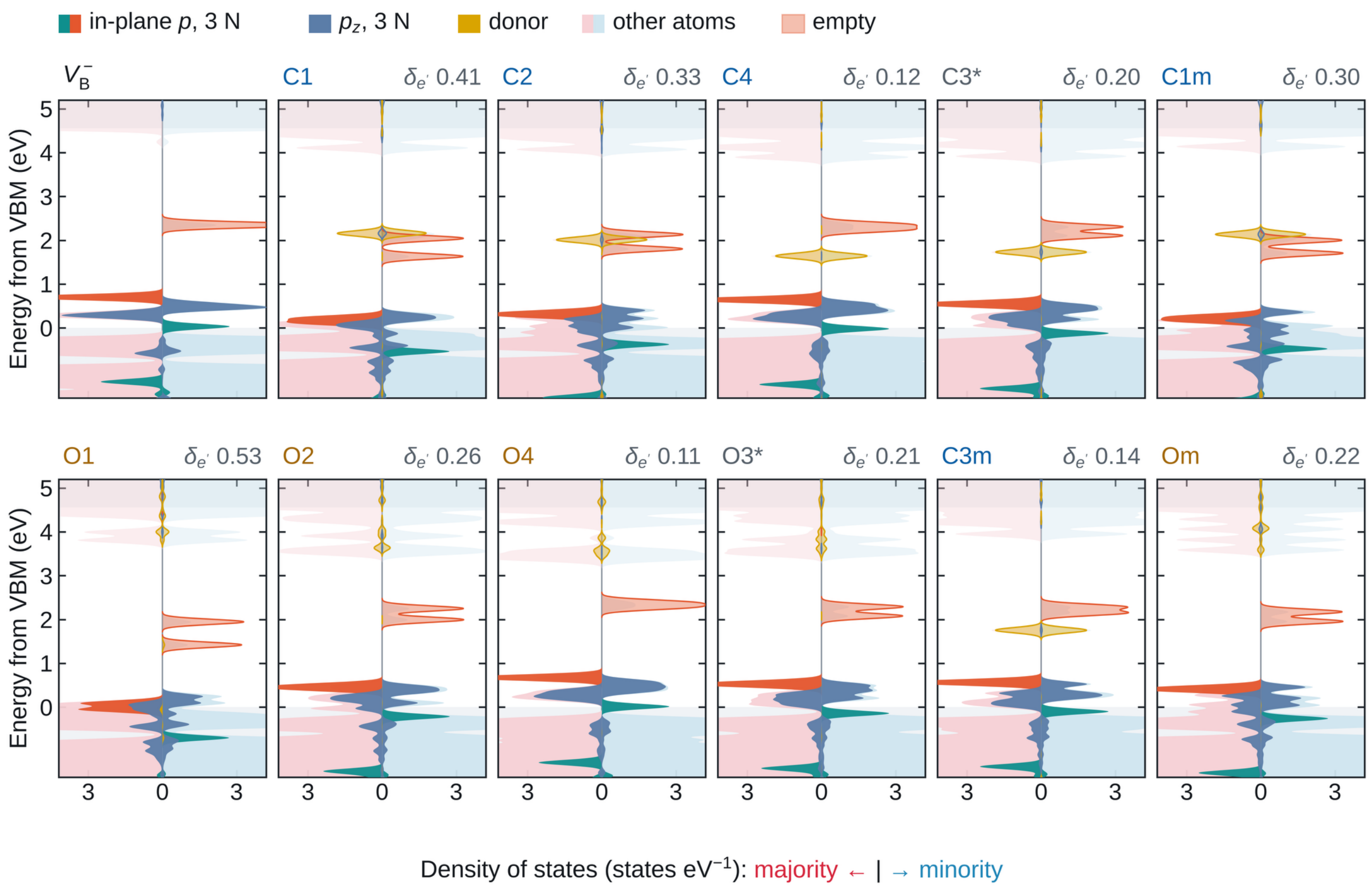

**Figure S6.** Local densities of states at $\Gamma$ of $V_B^-$ and of the eleven complexes in the frozen lattice ($G_0$; PBE, $8 \times 8$): levels broadened by Gaussians of 0.06 eV and projected on the in-plane $p$ ($a_1'$ teal, $e'$ coral) and $p_z$ (slate blue) orbitals of the three vacancy nitrogens, the donor (gold) and all other atoms (tinted, pale crimson majority, pale cerulean minority); light tints, empty states; shading, host bands of $V_B^-$; calculations aligned at $\Gamma$ on their highest occupied host-like level. $\delta_{e'}$, splitting of the empty minority $e'$ pair (eV). Every donor splits the hole doublet and keeps its own level empty; carbon levels lie near the doublet, oxygen levels higher.